\documentclass[pra,aps,superscriptaddress,twocolumn,floats,floatfix,nobibnotes]{revtex4-1}

\usepackage{natbib}
\usepackage{fullpage}

\usepackage{titlesec}
\usepackage{xcolor}

\PassOptionsToPackage{hyphens}{url}
\usepackage{hyperref}
\usepackage{etoolbox}

\usepackage{graphicx}

\usepackage{latexsym}
\usepackage{textcomp}
\usepackage{longtable}

\usepackage{array,multirow}
\usepackage{amsfonts,amsmath,amssymb, mathtools, dsfont}
\usepackage{bm}
\usepackage{braket}
\providecommand\citet{\cite}
\providecommand\citep{\cite}

\newif\iflatexml\latexmlfalse

\AtBeginDocument{\DeclareGraphicsExtensions{.pdf,.PDF,.eps,.EPS,.png,.PNG,.tif,.TIF,.jpg,.JPG,.jpeg,.JPEG}}

\usepackage[utf8]{inputenc}
\usepackage[english]{babel}

\newcommand{\eref}[1]{Eq.~\eqref{#1}}
\newcommand{\esref}[1]{Eqs.~\eqref{#1}}
\newcommand{\fref}[1]{Fig.~\ref{#1}}
\newcommand{\fsref}[1]{Figs.~\ref{#1}}
\newcommand{\Fref}[1]{Figure~\ref{#1}}
\newcommand{\sref}[1]{Sec.~\ref{#1}}
\newcommand{\aref}[1]{App.~\ref{#1}}

\newcommand{\rmmath}[1]{\mathrm{#1}}

\def \i {\mathrm{i}}

\def\XXint#1#2#3{{\setbox0=\hbox{$#1{#2#3}{\int}$}
     \vcenter{\hbox{$#2#3$}}\kern-.5\wd0}}

\newcommand{\Cq}{\mathcal{C}_\rmmath{q}}
\newcommand{\EPR}{\Delta_\rmmath{EPR}}
\usepackage{times}
\newcommand{\GammaRO}{\Gamma_\rmmath{ro}}

\definecolor{green}{rgb}{0.01, 0.75, 0.24}
\definecolor{red}{rgb}{1.0, 0.13, 0.32}

\begin{document}

\title{Stationary optomechanical entanglement between a mechanical oscillator and its measurement apparatus}

\author{C. Gut}
\altaffiliation[These authors contributed equally to this work.]{}
\affiliation{Vienna Center for Quantum Science and Technology (VCQ), Faculty of Physics,
University of Vienna, Boltzmanngasse 5, A-1090 Vienna, Austria}
\affiliation{Institute for Theoretical Physics and Institute for Gravitational Physics (Albert-Einstein-Institute), Leibniz University Hannover, Appelstrasse 2, 30167 Hannover, Germany}

\author{K. Winkler}
\altaffiliation[These authors contributed equally to this work.]{}
\affiliation{Vienna Center for Quantum Science and Technology (VCQ), Faculty of Physics,
University of Vienna, Boltzmanngasse 5, A-1090 Vienna, Austria}

\author{J. Hoelscher-Obermaier}
\affiliation{Vienna Center for Quantum Science and Technology (VCQ), Faculty of Physics,
University of Vienna, Boltzmanngasse 5, A-1090 Vienna, Austria}

\author{S. G. Hofer}
\affiliation{Vienna Center for Quantum Science and Technology (VCQ), Faculty of Physics,
University of Vienna, Boltzmanngasse 5, A-1090 Vienna, Austria}
\affiliation{Institute for Theoretical Physics and Institute for Gravitational Physics (Albert-Einstein-Institute), Leibniz University Hannover, Appelstrasse 2, 30167 Hannover, Germany}

\author{R. Moghadas Nia}
\affiliation{Vienna Center for Quantum Science and Technology (VCQ), Faculty of Physics,
University of Vienna, Boltzmanngasse 5, A-1090 Vienna, Austria}

\author{N. Walk}
\affiliation{Dahlem Center for Complex Quantum Systems, Physics Department, Freie Universit\"{a}t Berlin, Germany}

\author{A. Steffens}
\affiliation{Dahlem Center for Complex Quantum Systems, Physics Department, Freie Universit\"{a}t Berlin, Germany}

\author{J. Eisert}
\affiliation{Dahlem Center for Complex Quantum Systems, Physics Department, Freie Universit\"{a}t Berlin, Germany}

\author{W. Wieczorek}
\altaffiliation[Present address: ]{Department of Microtechnology and Nanoscience (MC2), Chalmers University of Technology, Kemiv\"agen 9, G\"oteborg, Sweden}
\affiliation{Vienna Center for Quantum Science and Technology (VCQ), Faculty of Physics,
University of Vienna, Boltzmanngasse 5, A-1090 Vienna, Austria}

\author{J. A. Slater}
\altaffiliation[Present address: ]{QuTech, Delft University of Technology, Lorentzweg 1, 2628 CJ Delft, The Netherlands}
\affiliation{Vienna Center for Quantum Science and Technology (VCQ), Faculty of Physics,
University of Vienna, Boltzmanngasse 5, A-1090 Vienna, Austria}

\author{M. Aspelmeyer}
\affiliation{Vienna Center for Quantum Science and Technology (VCQ), Faculty of Physics,
University of Vienna, Boltzmanngasse 5, A-1090 Vienna, Austria}
\affiliation{Institute for Quantum Optics and Quantum Information (IQOQI), Austrian Academy of Sciences, Boltzmanngasse 3, A-1090 Vienna, Austria}

\author{K. Hammerer}
\affiliation{Institute for Theoretical Physics and Institute for Gravitational Physics (Albert-Einstein-Institute), Leibniz University Hannover, Appelstrasse 2, 30167 Hannover, Germany}

\begin{abstract}
We provide an argument to infer stationary entanglement between light and a mechanical oscillator based on continuous measurement of light only. We propose an experimentally realizable scheme involving an optomechanical cavity driven by a resonant, continuous-wave field operating in the non-sideband-resolved regime. This corresponds to the conventional configuration of an optomechanical position or force sensor. We show analytically that entanglement between the mechanical oscillator and the output field of the optomechanical cavity can be inferred from the measurement of squeezing in (generalized) Einstein--Podolski--Rosen quadratures of suitable temporal modes of the stationary light field. Squeezing can reach levels of up to 50\% of noise reduction below shot noise in the limit of large quantum cooperativity. Remarkably, entanglement persists even in the opposite limit of small cooperativity. Viewing the optomechanical device as a position sensor, entanglement between mechanics and light is an instance of object--apparatus entanglement predicted by quantum measurement theory.
\end{abstract}

\date\today
\maketitle

\section{INTRODUCTION}
Experiments in optomechanics now operate routinely in a regime in which effects predicted by quantum theory can be observed. This includes mechanical oscillators cooled to their ground state of center-of-mass motion \cite{teufel_sideband_2011, Painter_2011}, ponderomotive squeezing \cite{Stamper-Kurn_2012, Safavi-Naeini_2013, Purdy2013a}, entanglement between different light tones \cite{Barzanjeh_2019, Schliesser_2020}, entanglement between different mechanical oscillators \cite{Ockeloen-Korppi2018, Riedinger2018}, measurement back-action \cite{Purdy2013}, back-action evasion \cite{Schwab_2016, Teufel_2017, Shomroni2019} and optomechanical entanglement, that is, entanglement between the mechanical oscillator and light, in a pulsed regime \cite{Palomaki_2013, Groeblacher_2018}.

It has been a long-standing prediction by Genes \emph{et al.}~\cite{Genes_2008} that optomechanical entanglement persists under certain conditions in steady state under continuous-wave drive, see Ref.~\cite{Genes_2009} for a comprehensive review. As explained in Ref.~\cite{Hofer_2011}, the experimental verification of entanglement between light and the mechanical oscillator is very challenging because the state of the latter is not accessible directly. The quantum state of the oscillator can, in principle, be reconstructed from measurements on light and suitable post-processing, as described in Refs.~\cite{Miao2, Miao3, Miao4} and demonstrated in Ref.~\cite{rossi_observing_2019}. In particular, Miao \emph{et al.}~\cite{Miao3} have discussed how stationary optomechanical entanglement can be inferred from such a complete quantum state tomography relying on temporally ordered preparation and verification steps. This approach is challenging in practice because it relies on an accurate characterization of all system parameters and back-action evading measurements. Moreover, a complete reconstruction, especially in the case of pure, entangled quantum states, is complicated by effects of finite measurement statistics, as we will show in the following. To date, it is still an open task to verify experimentally stationary optomechanical entanglement.

This work proposes an alternative approach that avoids a complete quantum state reconstruction. It is based on an argument that allows to infer optomechanical entanglement from the observation of entanglement between successive, non-overlapping light modes. Such an inference is possible under the assumption that the temporal modes of the light and the mechanical oscillator were not correlated initially, i.e., before the optomechanical interaction. The argument is analogous -- and in fact generalizes -- the one in the demonstration of pulsed optomechanical entanglement \cite{Palomaki_2013,Hofer_2011}. The first aspect of our work presents and proves this central argument.

The second aspect of this work describes a scheme to demonstrate experimentally stationary optomechanical entanglement, making use of the argument. The scheme applies to cavity optomechanical systems with large mechanical quality factor and large cavity linewidth as compared to the mechanical frequency (sideband unresolved regime). The systems must be in their steady state and driven by a continuous laser that is resonant with the cavity. Importantly, the drive is strictly constant (no modulations). The scheme is based on the measurement of an entanglement witness on suitably chosen temporally ordered modes of the light escaping the cavity -- the argument allows to infer optomechanical entanglement from witnessing entangled light modes. If the dynamics are stationary, the state is invariant under translation in time -- and so is the entanglement. We emphasize that the time-ordered light modes can be extracted from measured data in post-processing. The formulation of this scheme is simple enough that we could study it analytically and prove that it detects entanglement in the form of squeezing of Einstein–Podolski–Rosen (EPR) modes of light, or suitable generalizations thereof. We predict EPR-squeezing to scale inversely proportional to the quantum cooperativity (Ref.~\cite{Miao1} had found a similar behavior) and asymptotically reach 50\% of noise reduction below the shot noise level. This leaves a safe margin for experimental imperfections, such as finite detection efficiency. As in Ref.~\cite{Miao1}, we find that optomechanical entanglement persists even for a quantum cooperativity below unity, albeit only to a small extent.

The operating regime of the scheme we propose corresponds to the generic configuration of an optomechanical position or force sensor. In a continuous position sensor the measured object (the mechanical oscillator) and the measurement apparatus (the light field) share entanglement. This is the central idea of the quantum mechanical measurement theory \cite{VonNeumann, Zurek_81, Zurek_03}: the process underlying a physical measurement is ultimately an entangling interaction between object and the measuring apparatus (observer). This applies regardless of whether the measurement is performed at high quantum cooperativity, for which it is limited by measurement back-action, or at low cooperativity, where it is limited by shot noise or thermal noise. Under normal conditions, it is virtually impossible to detect the entanglement between object and apparatus as it usually involves an uncontrollable variety of environmental degrees of freedom due to the amplification associated with the measurement. It is intriguing to see that with an optical--mechanical sensor it is possible to capture this entanglement in a feasible measurement and thus shift the Heisenberg--von Neumann cut between object and observer.

This article is organized as follows: in \sref{sec_elementsOM} we introduce necessary elements of optomechanics. Our argument and scheme to detect optomechanical entanglement are developed in \sref{sec_entVerif}. We present our results and predictions regarding detectable signatures of optomechanical entanglement in \sref{sec_results}.

\section{ELEMENTS OF OPTOMECHANICS}\label{sec_elementsOM}

This section describes, concisely, the linearized cavity optomechanical model we rely on in the following. It is a well studied theory and many experiments demonstrated that it describes accurately diverse optomechanical devices operating in a wide range of parameter regimes; see the review on optomechanics \cite{OM_Review2014} for a thorough presentation.

We consider a standard optomechanical system \cite{OM_Review2014} comprising a single mechanical (oscillatory) mode interacting with a single light mode of a cavity. Mechanical and light fields are bosonic, described by two pairs of dimensionless Hermitian operators, $x_\rmmath{m}$/$p_\rmmath{m}$ and $x_\rmmath{c}$/$p_\rmmath{c}$, referred to as position/momentum and amplitude/phase quadratures of the mechanical and cavity mode, respectively. These operators obey equal-time canonical commutation relations,
\begin{align}
    [x_\alpha(t),p_\beta(t)] &= \i \delta_{\alpha, \beta},
\end{align}
where $\alpha,\beta=\rmmath{m,c}$ and $\delta_\alpha, \beta$ is the Kronecker delta symbol. Throughout this work $\hbar = 1$. We will also use ladder operators $a_\alpha$ and $a_\alpha^\dagger$ defined by
\begin{equation}\label{eq_quadToLadder}
    a_\alpha=(x_\alpha + \i p_\alpha)/\sqrt{2}
\end{equation}
obeying $[a_\alpha(t),a_\beta^\dagger(t)] = \delta_{\alpha, \beta}$.

In a frame rotating at the frequency $\omega_\rmmath{d}$ of the field driving the optomechanical cavity, the Hamiltonian of the linearized dynamics is given by \cite{OM_Review2014}
\begin{equation}\label{eq_Hamiltonian}
H = \omega_\rmmath{m}a_\rmmath{m}^\dagger a_\rmmath{m} -\delta a_\rmmath{c}^\dagger a_\rmmath{c} +g(a_\rmmath{m}^\dagger a_\rmmath{c}^\dagger + a_\rmmath{m} a_\rmmath{c}^\dagger + \text{H.c.}).
\end{equation}
We denote the oscillation frequency of the mechanics  by $\omega_\rmmath{m}$, $\delta \approx \omega_\rmmath{d} -\omega_\rmmath{c}$ is the detuning of the drive from cavity resonance, and $g$ is the coupling strength (depending on the drive power). The interaction contains two processes that create sidebands in the cavity field: the down-conversion of a phonon and a photon at a lower energy than the drive (that is, Stokes scattering to sideband frequency $-\omega_\rmmath{m}$), and the state-swap of a phonon onto a photon at higher energy than the drive (that is, anti-Stokes scattering to sideband frequency $+\omega_\rmmath{m}$).

For completeness (this is a technical note), we mention here that the non-linear optomechanical Hamiltonian can be written in terms of a Kerr interaction in the cavity \cite{Rabl2011, Nunnenkamp2011}. Subsequent linearization gives a single-mode (degenerate) down-conversion interaction term:  $\left[a_\rmmath{c}^2+(a_\rmmath{c}^\dagger)^2 \right] g^2/\omega_\rmmath{m}$. It is possible to show that such a process is not sufficient to generate the entanglement we propose to reveal with our scheme (described in the coming \sref{sec_entVerif}). To do so, apply the mode functions \eref{eq_modeFunct_time} to the two-time correlators of a degenerate parametric down-converter \cite[ch. 10]{Zoller_Gardiner2010} and evaluate the integrals -- this is an application of the formalism developed in \aref{sec_DerivEPRvarGeneral} -- then check that our entanglement criterion \eref{eq_DuanCrit} is never violated this way. Therefore, the linearized model \eqref{eq_Hamiltonian} is the relevant one for our purpose.

The cavity mode is coupled to the free electromagnetic field outside the cavity. The ladder operators describing this one-dimensional field, $a_\rmmath{in}(t)$ and $a_\rmmath{in}^\dagger(t)$, have the following Markovian correlation functions
\begin{subequations}\label{eq_lightBath_correl}
\begin{align}
    \langle a_\rmmath{in}(t) \rangle &= 0, & \langle a_\rmmath{in}^\dagger(t) \rangle &= 0,\\
    \langle a_\rmmath{in}^\dagger(t) a_\rmmath{in}(t') \rangle &= 0, & \langle a_\rmmath{in}(t) a_\rmmath{in}^\dagger(t') \rangle &= \delta(t-t').
\end{align}
\end{subequations}
We neglect here thermal occupation numbers for optical frequencies.

The thermal bath  of the mechanical oscillator is modeled by
quantum Brownian motion damping with the Hermitian noise operator $\xi$ \cite{Caldeira_Leggett_83, Zoller_Gardiner2010}. At high temperatures of the mechanical bath $n_\mathrm{th} \approx k_\rmmath{B}T_\rmmath{bath}/\hbar \omega_\rmmath{m}\gg1$ and/or large mechanical quality factor $Q=\omega_\rmmath{m}/\gamma_\rmmath{m} \gg 1$ its correlators are approximately Markovian, reflected by
\begin{subequations}\label{eq_mechBath_correl}
\begin{align}
    \langle \xi(t)\rangle &= 0,\\
    \langle \xi(t) \xi(t') + \xi(t') \xi(t) \rangle &\approx (2 n_\mathrm{th}+1) \delta(t-t').
\end{align}
\end{subequations}

Including cavity (power) decay at rate $\kappa$ and viscous damping of the oscillator at rate $\gamma_\mathrm{m}$, the open-system dynamics is described by the quantum Langevin equations (QLE) \cite{OM_Review2014}\footnote{The QLE can be either expressed with the Brownian noise affecting the mechanical momentum quadrature only, as in \esref{eq_QLEtime}, or with the Brownian noise affecting both mechanical quadratures symmetrically. We performed the exact numerical computation based on the QLE with symmetric and asymmetric Brownian noise and we found that all the results presented in this work are the same in both cases, up to a few parts per million.}
\begin{subequations}\label{eq_QLEtime}
\begin{align}
    \dot{x}_\rmmath{m}&= \omega_\rmmath{m} p_\rmmath{m},\\
    \dot{p}_\rmmath{m}&= -\gamma_\rmmath{m} p_\rmmath{m} -\omega_\rmmath{m} x_\rmmath{m} -2gx_\rmmath{c} +\sqrt{2\gamma_\rmmath{m}} \xi\label{eq_QLEtime-2}, \\
    \dot{x}_\rmmath{c}&=-\delta p_\rmmath{c} -\frac{\kappa}{2}x_\rmmath{c} +\sqrt{\kappa}x_\rmmath{in} \label{eq_QLEtime-3}, \\
     \dot{p}_\rmmath{c}&=\delta x_\rmmath{c} -\frac{\kappa}{2}p_\rmmath{c} -2gx_\rmmath{m} +\sqrt{\kappa}p_\rmmath{in}.
\end{align}
\end{subequations}

In the following, we consider the special case of a resonantly driven cavity, $\delta=0$, which corresponds to the standard configuration of an optomechanical position or force sensor. The generalisation to nonzero detuning is straightforward, but results in significantly more involved analytical expressions that we will not reproduce here. In \esref{eq_QLEtime}, $x_\rmmath{in}$ and $p_\rmmath{in}$ correspond to shot noise because we work in a suitably displaced frame where mean amplitudes were shifted to zero, see Ref.~\cite{OM_Review2014} for details.

Our study always assumes stable dynamics such that the steady-state is reached eventually \footnote{The stability is most easily assessed by the Routh--Hurwitz criterion. Whenever the detuning is zero the dynamics is unconditionally stable \cite{Thesis_Hofer}.}. Equations \eqref{eq_QLEtime} can be solved easily in Fourier space \cite{Genes_2008}; see \aref{sec_SolQLE} for the Fourier transform conventions we have
adopted. The quadratures of the light emitted by the cavity are given by input--output relations \cite{Gardiner_Collet_85} and take the form
\begin{subequations}\label{eq_outputQuad}
\begin{align}
    x_\rmmath{out}(\omega) &= \mathcal{S}(\omega) x_\rmmath{in}(\omega),\\
    p_\rmmath{out}(\omega) &= \mathcal{S}(\omega) p_\rmmath{in}(\omega) + 4g^2 \chi_\rmmath{opt}^2(\omega) \chi_\rmmath{m}(\omega) x_\rmmath{in}(\omega) \nonumber\\
    &\quad - 2g \sqrt{2\gamma_\rmmath{m}} \chi_\rmmath{opt}(\omega) \chi_\rmmath{m}(\omega) \xi (\omega).\label{eq_outputQuad_pout}
\end{align}
\end{subequations}
We have introduced here the mechanical and the optical susceptibilities
\begin{align}
\chi_\rmmath{m}(\omega) &\coloneqq \frac{\omega_\rmmath{m}}{\omega_\rmmath{m}^2 - \omega^2 -\i \omega \gamma_\rmmath{m}}, \label{eq_MecSusc}\\
\chi_\rmmath{opt}(\omega) &\coloneqq \frac{\sqrt{\kappa}}{\frac{\kappa}{2} -\i \omega}, \label{eq_CavSusc}
\end{align}
and the reflection phase
\begin{align}
 \mathcal{S}(\omega) \coloneqq \frac{\frac{\kappa}{2} +\i \omega}{\frac{\kappa}{2} -\i \omega}. \label{eq_ReflPhase}
\end{align}

We assume $Q = \omega_\rmmath{m}/\gamma_\rmmath{m} \gg1$, which is a typical feature in micro-optomechanical setups. This allows to approximate the poles of the mechanical susceptibility, \eref{eq_MecSusc}, by $\omega_\pm \approx \pm \omega_\rmmath{m} - \i \gamma_\rmmath{m}/2$, such that
\begin{subequations}\label{eq_approximatedFunctions}
\begin{align}
    \chi_\rmmath{m}(\omega) &\approx \frac{1}{2}\left( \frac{1}{\omega - \omega_-} - \frac{1}{\omega - \omega_+} \right),\label{eq_MecSuscApprox} \\
    |\chi_\rmmath{m}(\omega)|^2 &\approx \frac{1}{4}\left( \frac{1}{|\omega - \omega_-|^2} + \frac{1}{|\omega - \omega_+|^2} \right).\label{eq_MecSuscApprox_modSquare}
\end{align}
Both are strongly peaked at $\pm \omega_\rmmath{m}$. Close to these frequencies and in the  sideband unresolved regime ($\kappa\gg \omega_\rmmath{m}$) we approximate \esref{eq_CavSusc} and \eqref{eq_ReflPhase} by, respectively,
\begin{align}\label{eq:approx_chiopt}
    \chi_\rmmath{opt}(\omega) &\approx \frac{2}{\sqrt{\kappa}}, &
    \mathcal{S}(\omega) &\approx 1.
\end{align}
\end{subequations}

 In the limit of these approximations, the characteristic response time $1/\kappa$ of the intra-cavity field is the shortest time scale of the system, such that the intra-cavity field is adiabatically eliminated from the dynamics. In other words, the mechanical oscillator is effectively directly coupled to the output field, without any spectral filtering due to the cavity, cf. \eref{eq:approx_chiopt}. In this limit, one can rewrite \eref{eq_outputQuad_pout} as
\begin{align}
    p_\rmmath{out}(\omega) &= \mathcal{S}(\omega) p_\rmmath{in}(\omega) + 4 \GammaRO \chi_\rmmath{m}(\omega) x_\rmmath{in}(\omega) \nonumber\\
    &\quad - 2\sqrt{2\gamma_\rmmath{m}\GammaRO} \chi_\rmmath{m}(\omega) \xi (\omega),\label{eq_outputQuad_pout1}
\end{align}
where we have defined the readout rate
\begin{align}\label{eq_GammaRO}
{\GammaRO} \coloneqq \frac{4g^2}{\kappa}.
\end{align}
The first term of $p_\rmmath{out}$ in \eref{eq_outputQuad_pout1} is the contribution of shot noise reflected by the cavity. The second term is the contribution from shot noise driving the mechanical motion and being transferred back to the light via the interaction -- this effect is called \emph{back-action}. The last term is the contribution of the thermal fluctuations from the mechanical bath, mapped onto the light via the optomechanical interaction. In the context of position or force sensing, this term constitutes the signal to be determined from a measurement of the phase quadrature $p_\rmmath{out}$. The relative size of the readout term to the thermal noise in \eref{eq_outputQuad_pout1} is the so-called quantum cooperativity
\begin{align}\label{eq_cooperativity}
\Cq \coloneqq \frac{4 g^2}{\kappa \gamma_\rmmath{m}(n_\rmmath{th}+1)}\approx \frac{\GammaRO}{\Gamma_\rmmath{th}}.
\end{align}
The last approximation holds in the high temperature limit, and we defined the thermal decoherence rate of the mechanical oscillator
\begin{align}\label{eq_Gammath}
  \Gamma_\rmmath{th}  \coloneqq \gamma_\rmmath{m}(n_\rmmath{th}+1/2),
\end{align}
which sets the scale of the thermal force in \eref{eq_QLEtime-2}.

At a fundamental level, the transduction of information on position or force is associated with the generation of correlations, or even quantum entanglement, between the observed object (here, the mechanical oscillator) and the measurement apparatus (the light field). In the next section we will explain how stationary optomechanical entanglement can be proven unambiguously based solely on measurements of the light field.

\section{VERIFICATION OF OPTOMECHANICAL ENTANGLEMENT} \label{sec_entVerif}

\subsection{Inference of optomechanical entanglement from measurement of light} \label{sec_argument}
\begin{figure}
\includegraphics[width=\columnwidth]{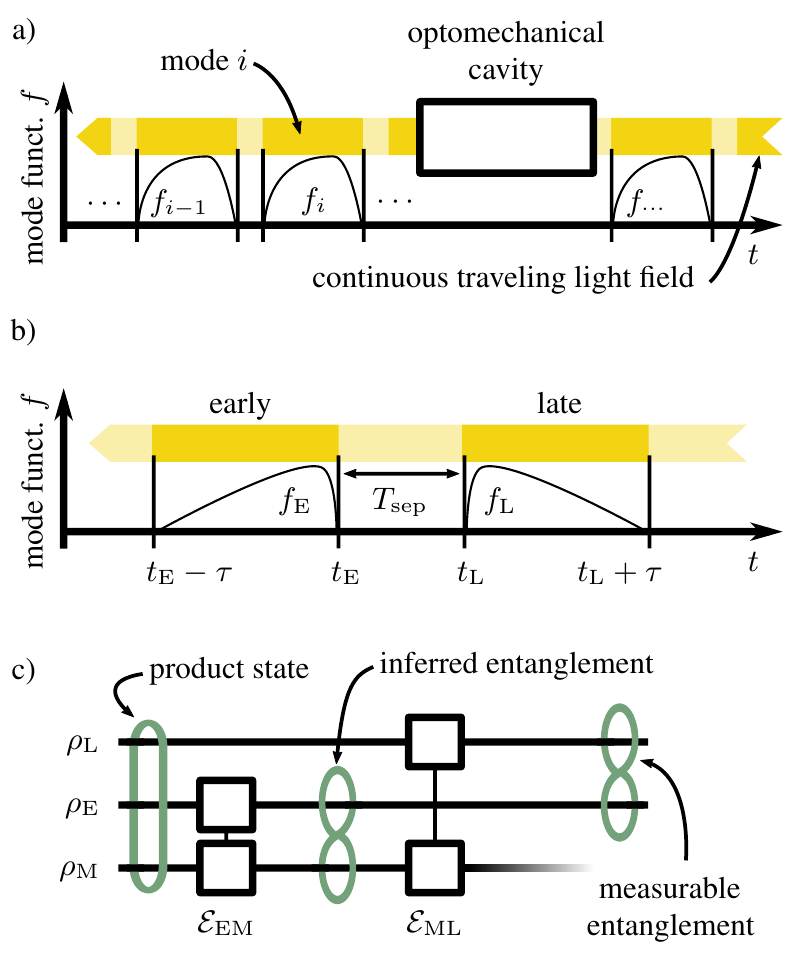}
\caption{a) The continuous light field (traveling to the left because the time axis points to the right) can be decomposed (abstractly) into temporal modes via an abstract modulation by mode functions $f_i$. The shaded area is a temporal separation between the modes. b) Two temporal modes are considered: an early pulse~(E) and a late pulse~(L) defined by the mode functions $f_\mathrm{E}$ and $f_\mathrm{L}$. c) The quantum circuit corresponding to the dynamics. The mechanical oscillator~(M), the early and the late pulses are initially in an uncorrelated state. M interacts sequentially with E and then L according to the quantum channels $\mathcal{E}_\text{EM}$ and $\mathcal{E}_\text{ML}$, respectively.} \label{fig_quantumCircuit}
\end{figure}

 The continuous traveling light field entering and escaping an optomechanical cavity can be decomposed into pulses, i.e., a sequence of non-overlapping temporal modes. A temporal mode labeled $i$ is defined by a mode function $f_i$, see \fref{fig_quantumCircuit} a). We restrict ourselves to two consecutive modes of such a decomposition: an early pulse~(E) and a late pulse~(L). They have support on time intervals $[t_\mathrm{E}-\tau,t_\mathrm{E}]$ and $[t_\mathrm{L},t_\mathrm{L}+\tau]$, respectively, with $t_\mathrm{E}<t_\mathrm{L}$, $T_\mathrm{sep} \coloneqq t_\mathrm{L}-t_\mathrm{E}$ their separation, and $\tau>0$ their duration, see \fref{fig_quantumCircuit} b). In this picture, the mechanical oscillator (M) interacts first with the early pulse, according to a quantum channel (i.e., a completely positive map) $\mathcal{E}_\mathrm{EM}$, and then with the late pulse, according to $\mathcal{E}_\mathrm{ML}$. This temporal ordering of the dynamics is illustrated by the quantum circuit in \fref{fig_quantumCircuit} c). We assume that the two pulses and the mechanical oscillator are initially in a product state $\rho_\mathrm{EML}^\mathrm{init}$. The final state of the three systems is $\rho^\mathrm{fin}_\mathrm{EML}= (\mathds{1}_\mathrm{E}\otimes\mathcal{E}_\mathrm{ML})(\mathcal{E}_\mathrm{EM}\otimes \mathds{1}_\mathrm{L}) \rho_\mathrm{EML}^\mathrm{init}$. If the reduced state of the mechanical oscillator and the early pulse after the first channel $\rho_\rmmath{EM}=\mathrm{tr}_\mathrm{L}(\mathcal{E}_\mathrm{EM}\rho_\mathrm{EML}^\mathrm{init})$ is separable, then the final state of early and late pulses $\rho_\mathrm{EL}=\mathrm{tr}_\mathrm{M} (\rho^\mathrm{fin}_\mathrm{EML})$ is separable too. Conversely, if $\rho_\mathrm{EL}$ is entangled, $\rho_\mathrm{EM}$ must have been entangled. Thus, the measurement of entanglement between early and late pulses implies optomechanical entanglement. A detailed proof of this intuitive but non-trivial statement is presented in \aref{sec_ArgumentDetails}.

Based on this reasoning, entanglement between a mechanical resonator and a travelling wave pulse of a microwave field had been demonstrated in Ref.~\cite{Palomaki_2013}, as outlined in Ref.~\cite{Hofer_2011}. Optomechanical entanglement had been deduced from the measurement of entanglement between two consecutive microwave pulses that originated from sequential interactions with the mechanical oscillator. In the scheme of Refs.~\cite{Hofer_2011,Palomaki_2013}, the two processes $\mathcal{E}_\rmmath{EM}$ and $\mathcal{E}_\rmmath{ML}$ are different in nature, and constructed to optimize entanglement generation in $\mathcal{E}_\rmmath{EM}$ and state readout during $\mathcal{E}_\rmmath{ML}$. This requires, in particular, to work in the resolved-sideband limit ($\kappa/\omega_\mathrm{m}<1$), and to set the driving field blue-detuned ($\delta=\omega_\mathrm{m}$) in the first (E--M) interaction and red-detuned ($\delta=-\omega_\mathrm{m}$) during the second (M--L) interaction. In both processes the applied fields are pulsed and steer the optomechanical system through non-stationary, non-equilibrium states.

Our argument, at the beginning of this section, has two premises that the system must fulfill: time ordering of the interactions and no initial correlations. In principle, no additional knowledge about the system is needed to use the argument to relate the presence of entanglement in the light to the presence of optomechanical entanglement. The argument makes a sufficient inference: if the temporal modes are entangled, then there is optomechanical entanglement; if they are not, no conclusion can be drawn. This is in contrast with the method proposed in Refs.~\cite{Miao2, Miao3, Miao4} that can characterize completely the optomechanical entanglement (full state tomography), based on extended and precise knowledge about the system. The trade-off between both methods is the amount of knowledge about the system necessary to obtain information about the oscillator. In this sense, both approaches are complementary.

 The requirement that the three parties involved are initially uncorrelated excludes the hypothetical scenario of Refs.~\cite{Cubitt_2003, Korolkova_08, Fedrizzi_2013, Peuntinger_2013, Vollmer_2013}. There, entanglement is distributed via a mediator, without the mediator ever becoming entangled (see end of \aref{sec_ArgumentDetails} for details). Since in real experiments the light might contain some degree of classical correlations, demonstrating entanglement between certain modes of light that were mediated by a mechanical oscillator \cite{Barzanjeh_2019,Schliesser_2020} does not allow to make unambiguous statements regarding genuine optomechanical entanglement.

Light in the cavity has a life-time (of the order $1/\kappa$) that effectively delays the propagation of an incoming pulse. To guarantee the strictly sequential interaction depicted in \fref{fig_quantumCircuit} c), the temporal separation $T_\mathrm{sep}$ between the early and late pulses must be lager than $1/\kappa$, so that the pulses interact with the mechanics one after the other. In this work, we will focus on the regime where $1/\kappa$ is the shortest time scale; this choice will be motivated (physically) in \sref{sec_tempModes} by our choice of mode functions $f_\rmmath{E}$ and $f_\rmmath{L}$. However, in principle, it is not excluded that optomechanical entanglement can be demonstrated in systems with relatively narrow linewidth cavities.

In the following, we will use the logic of the argument presented above to show that optomechanical entanglement can be verified in the regime of continuous driving of an optomechanical cavity by a constant laser with fixed detuning, such that the mechanical oscillator and the output field of the cavity are in a stationary state. The two successive temporal pulses are extracted in post-processing from the continuous measurement on the output field. Because the overall state of the mechanical oscillator and the output field is stationary, it does not matter which intervals $[t_\mathrm{E}-\tau,t_\mathrm{E}]$ and $[t_\mathrm{L},t_\mathrm{L}+\tau]$ are extracted from the stationary homodyne current: any of its properties, such as entanglement of early and late pulses, will depend only on the pulse length $\tau$ and separation $T_\rmmath{sep}$ between the pulses, but not on the particular instants of time $t_\mathrm{E}$ and $t_\mathrm{L}$. Formally we define annihilation operators corresponding to these temporally ordered modes by
\begin{subequations}\label{eq:pulses}
\begin{align}
    r_\rmmath{E} &\coloneqq \int_{t_\mathrm{E}-\tau}^{t_\mathrm{E}} \rmmath{d}t\, f_\rmmath{E}(t) a_\rmmath{out}(t),\\
    r_\rmmath{L} &\coloneqq \int_{t_\mathrm{L}}^{t_\mathrm{L}+\tau} \rmmath{d}t\, f_\rmmath{L}(t) a_\rmmath{out}(t),
\end{align}
\end{subequations}
where $f_\rmmath{E}(t)$ and $f_\rmmath{L}(t)$ are the temporal mode functions of the early and late modes, respectively. The properties of $a_\rmmath{out}$ are determined by \esref{eq_outputQuad}, which describe the stationary state of light in conjunction with the properties of the ongoing noise processes \esref{eq_lightBath_correl} and \eqref{eq_mechBath_correl}. Because the two modes E and L are ordered in time and interact sequentially with the mechanical oscillator, the scenario of \fref{fig_quantumCircuit}~c) applies.  Before the interaction with the mechanics, modes E and L are defined with respect to $a_\mathrm{in}$ in \eref{eq:pulses} (distinct modes of shot noise in our displaced frame), hence they are not correlated to the oscillator nor to each other. Therefore the argument applies and optomechanical entanglement is demonstrated through observing that the two light modes share entanglement.

It is instructive to compare this once again to the pulsed scheme of Refs.~\cite{Hofer_2011,Palomaki_2013}. There the properties of early and late pulses had to be inferred from an integration in time of the respective equations of motion, \esref{eq_QLEtime}, with different detuning $\delta=\pm\omega_\mathrm{m}$ for E and L. Here we can infer all properties of E and L from the stationary solutions, \esref{eq_outputQuad}, for a fixed detuning. In the pulsed scheme the description of the three modes E, L and M was essentially complete in the sense that the dynamics were designed so that no correlations to any other mode of the light field were established. Here, the restriction to the two pulses E and L is a massive simplification: in reality, the mechanical oscillator will exhibit correlations to more modes than E and L, and the stationary light field will exhibit a large variety of internal correlations, e.g., in the form of ponderomotive squeezing. Still, for demonstrating stationary entanglement of the mechanical oscillator and light, it  is sufficient to consider the three modes E, M and L.

For the following discussion it will be  mathematically convenient to formally allow for infinitely extended pulses, corresponding to $\tau\rightarrow\infty$, whose pulse envelopes $f_\rmmath{E/L}(t)$ tend to zero for $t\rightarrow\mp\infty$, respectively (keeping the mode finite does not change the derivation). The characteristic scale at which the envelopes tend to zero defines an effective pulse length. Furthermore, because of stationarity, we can set arbitrarily $t_\mathrm{E}$ to $-T_\rmmath{sep}/2$ and $t_\mathrm{L}$ to $T_\rmmath{sep}/2$ without loss of generality. We write the mode operators $r_{i}$, for $i=\rmmath{E,L}$, as
\begin{align}\label{eq_modeOp_time}
    r_{i} &= \int_{-\infty}^\infty \rmmath{d}t\, f_{i}(t) a_\rmmath{out}(t)
\end{align}
where $f_\rmmath{E}(t)$ has to be an anti-causal function with respect to $-T_\rmmath{sep}/2$ (i.e., $f_\rmmath{E}(t)= 0$ for $t>-T_\rmmath{sep}/2$) and $f_\rmmath{L}(t)$ has to be causal with respect to $T_\rmmath{sep}/2$ (i.e., $f_\rmmath{L}(t)= 0$ for $t<T_\rmmath{sep}/2$). The $r_i$ must fulfil bosonic commutation relations, $[r_i(t), r_j^\dagger(t)] = \delta_{i,j}$, which correspond to an orthonormalisation constraint for the mode functions
\begin{equation}\label{eq_orthoNormalisation}
        \int_{-\infty}^\infty\rmmath{d}t\, f_i(t) f_j^*(t) \overset{!}{=} \delta_{i,j}\ ;\quad i,j= \rmmath{E,L}.
\end{equation}
With \eref{eq_quadToLadder} we obtain the quadrature operators $x_i$, $p_i$ of the early and late modes of the output field.

In order to prove stationary optomechanical entanglement, we need to identify suitable mode functions and system parameters that result in an entangled state $\rho_\text{EL}$ of early and late pulses, which can be verified from a measurement of the output light field of the optomechanical cavity.

In the following \sref{sec_entCriterion} we motivate a choice of entanglement criterion. This choice specifies what information the measurement procedure must retrieve from the output field. In \sref{sec_tempModes} we motivate a particular class of temporal mode functions to process measurement data. This choice is based on our understanding of the stationary optomechanical dynamics and leads to choosing the particular parameter regime of unresolved sidebands and undetuned drive. These choices determine an implementable experimental scheme that we can study analytically; the results of the study are presented in \sref{sec_results}.

\subsection{Entanglement criteria}\label{sec_entCriterion}

We note first that the optomechanical dynamics is linear and the driving fields correspond to Gaussian white noise. This implies that the stationary state of the optomechanical system and the output field is Gaussian, and that the state of the two temporal modes in the output field, introduced in \eref{eq_modeOp_time}, is Gaussian too. First moments and covariances of canonical operators $x_i, p_i$ ($i=\rmmath{E,L}$) determine Gaussian states completely. As a consequence all entanglement properties are fully determined by the latter. Entanglement of the two-mode Gaussian state of E and L can be quantified by means of suitable entanglement measures \cite{Continuous,Adesso_2007,GaussianQuantumInfo} (such as the Gaussian entanglement of formation \cite{PhysRevA.69.052320} or the logarithmic negativity \cite{VidalNegativity,PlenioNegativity,PhD}), which can be calculated from the experimentally accessible covariances of the state $\rho_\mathrm{EL}$. What is more, such entanglement measures often provide bounds to non-Gaussian states for given covariances.

In order to proceed in a fully systematic way, one could try to optimize temporal mode functions in \eref{eq_modeOp_time} with respect to an entanglement measure and other system parameters (such as optomechanical coupling $g$, etc.), as in Ref.~\cite{Miao1}. However, due to the highly nonlinear dependence of entanglement measures on the state $\rho_\mathrm{EL}$ such an optimization seems  too challenging. Moreover, our argument that relates entanglement in the light field to optomechanical entanglement (presented at the beginning of \sref{sec_argument}) is a sufficient condition for entanglement and, on its own, does not allow to relate the amount of entanglement (characterized by a measure) shared by the light modes to the amount of entanglement between the oscillator and the light.

Instead, we use entanglement criteria that are linear in covariances, i.e., second moments of quadratures \cite{Duan_2000,Simon00,Hyllus_2006}.  Geometrically, these criteria specify separating hyperplanes from the convex set of covariances that are compatible with separable Gaussian states \cite{Hyllus_2006}. On the level of covariances, the geometry is that of an entanglement witness for quantum states. Since each test is linear in the covariances, each test defines an implementable and feasible measurement procedure. What is more, one can argue that a strong violation of such criteria is accompanied by a quantitative statement, in that several entanglement measures are lower bounded by quantitative violations of such entanglement tests \cite{Hyllus_2006,quant-ph/0607167, Audenaert06, Guehne}. Moreover,  we will use in \sref{sec_results} that, for an anticipated covariance, one can efficiently find the optimal test that best certifies the entanglement present in a Gaussian state \cite{Hyllus_2006}.

For our analytical consideration we will resort to a special case of such criteria, referred to as Duan's criterion~\cite{Duan_2000}: the so-called EPR-variance ($\EPR$)
\begin{equation}\label{eq_EPRvar_quad}
  \EPR \coloneqq \Delta \left(x_\text{E}+x_\rmmath{L}^\phi \right)+ \Delta \left(p_\text{E}-p_\rmmath{L}^\phi \right),
\end{equation}
where
\begin{align*}
  x_\rmmath{L}^\phi&=x_\rmmath{L} \,\cos\phi\, +p_\rmmath{L} \,\sin\phi, \\
  p_\rmmath{L}^\phi&=p_\rmmath{L} \,\cos\phi\, -x_\rmmath{L} \,\sin\phi.
\end{align*}
Our notation for the variances is
\begin{align}
\Delta A \coloneqq \langle A^2 \rangle_{\rho_\rmmath{EL}^\rmmath{init}} - \langle A \rangle_{\rho_\rmmath{EL}^\rmmath{init}}^2,
\end{align}
where the expectation values are taken with respect to the initial density operator $\rho_\rmmath{EL}^\rmmath{init} = \ket{0}\bra{0}_\rmmath{E} \otimes \ket{0}\bra{0}_\rmmath{L}$ (in our displaced frame). $\EPR$ quantifies the simultaneous correlations between pairs of quadratures of \emph{different modes}. For two-mode Gaussian states,
\begin{align}\label{eq_DuanCrit}
  \EPR <2
\end{align}
implies (only sufficient) that the state is entangled. In phase space, this corresponds to a reduction of the Gaussian state's variance (uncertainty) below the variance of a two-mode vacuum state (the coherent state that corresponds the shot noise of the temporal modes) along a particular direction in the plane $\left(x_\text{E}+x_\rmmath{L}^\phi \right)$ \& $\left(p_\text{E}-p_\rmmath{L}^\phi \right)$: this is a particular form of two-mode-squeezing that we call EPR-squeezing. In terms of ladder operators, the EPR-variance can be written as
\begin{multline}
  \EPR =\langle r_\rmmath{E} r_\rmmath{E}^\dagger + r_\rmmath{E}^\dagger r_\rmmath{E} + r_\rmmath{L} r_\rmmath{L}^\dagger + r_\rmmath{L}^\dagger r_\rmmath{L} \rangle\\
\quad+\left(e^{i\phi} \langle r_\rmmath{L} r_\rmmath{E} +r_\rmmath{E} r_\rmmath{L}\rangle + \rmmath{H.c.}\right). \label{eq_EPRvar_ladderOp}
\end{multline}
This relation can be readily evaluated from a record of homodyne measurements, along with appropriate post-processing in order to extract adequate (anti-)causal temporal modes. Again, the Duan criterion is in general not an optimal test, but can be optimized for a given Gaussian state.

\subsection{Temporal modes}\label{sec_tempModes}
In this section we address the question of how to choose the mode functions $f_\rmmath{E/L}$ in \eref{eq_modeOp_time} in order to achieve largest violation of the separability bound \eref{eq_DuanCrit}. The EPR-variance can be expressed as a quadratic form of the mode functions, which we show explicitly in \aref{sec_DerivEPRvarGeneral}, cf. \eref{eq_exactEPR}. Unfortunately the minimization does not map to a simple eigenvalue problem due to the required (anti-)causality and normalization of the mode functions.

\begin{figure}
\includegraphics[width=\columnwidth]{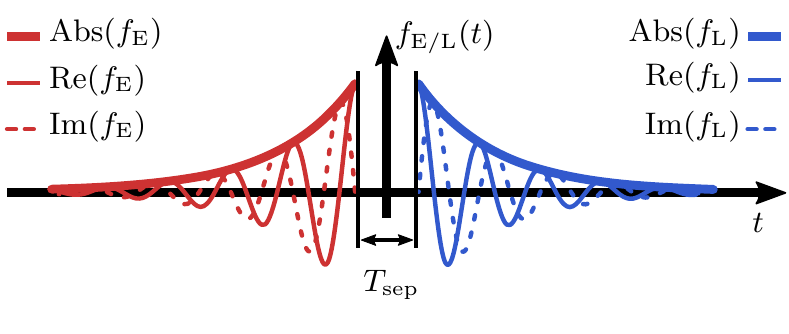}
\caption{Mode functions of \esref{eq_modeFunct_time}: they are temporally ordered and non-overlapping as required by the entanglement verification argument of \sref{sec_argument}. They are orthonormal as required by \eref{eq_orthoNormalisation}. And they are pulses to account for decoherence. The early pulse (in red) has a raising exponential time envelope and is resonant with the red sideband at $-\omega_\rmmath{m}$. The late pulse (in blue) has an exponentially decaying envelope in time and is resonant with the blue sideband at $+\omega_\rmmath{m}$.}\label{fig_modeFunct}
\end{figure}

We therefore proceed by choosing a suitable variational class of mode functions that we motivate as follows: in section \ref{sec_elementsOM}, we explained that the optomechanical unitary dynamics produce sidebands around the driving frequency at $\pm \omega_\rmmath{m}$ (in the frame rotating at the drive's frequency). The lower (red) sideband is produced by the down-conversion part (also termed two-mode-squeezing) of the interaction and it is known to produce EPR-like entanglement \cite{Genes_2008, Hofer_2011}: this sideband is entangled with the mechanics. The upper (blue) sideband is produced by the beam-splitter contribution of the interaction and corresponds to a coherent state-swap of the mechanical and cavity fields (preserving quantum correlations with the red sideband): this sideband encodes the state of the mechanics, which was entangled with the red sideband some moments before. The blue sideband is thus entangled with the light in the red sideband at previous times. If $\delta =0$ and if the cavity linewidth $\kappa$ is not much narrower than $\omega_\rmmath{m}$, then both sidebands will have significant spectral components in the measured output -- this is the reason for choosing the regime of unresolved sidebands and resonant drive. We choose early and late mode functions that capture the temporal order of these processes: the early mode extracts information from the red sideband and the late mode from the blue sideband. This is done with a demodulation of the detected signal by $e^{\pm\i \omega_\rmmath{m} t}$, which makes the corresponding spectral component resonant.

\begin{figure}
\includegraphics[width=\columnwidth]{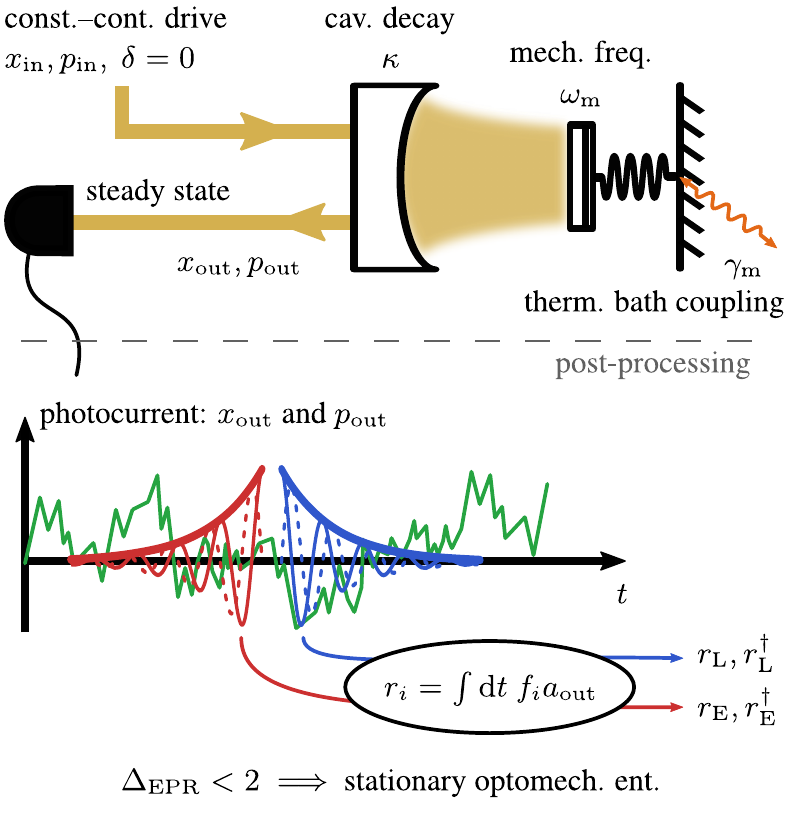}
\caption{Summary of the scheme we propose: consider a mechanical resonator that is well isolated from its thermal bath ($\omega_\mathrm{m}\gg \gamma_\mathrm{m}$), in a cavity not resolving sidebands ($\kappa \gg \omega_\mathrm{m}$), and driven in its steady state by a constant--continuous and resonant ($\delta=0$) laser ($x_\text{in}$, $p_\text{in}$). The output field ($x_\text{out}$, $p_\text{out}$) is continuously monitored, e.g., by homodyning. The corresponding measurement record is overlapped with appropriate mode functions $f_\text{E/L}$ [\esref{eq_modeOp_time}] to yield the ladder operators of the temporal modes ($r_\text{E/L}$, $r_\text{E/L}^\dagger$). Entanglement is tested with the EPR-variance \eref{eq_EPRvar_ladderOp}.} \label{fig_scheme}
\end{figure}

We expect the quantum correlations at different points in time to decrease with the duration separating them. This is because the incoherent coupling to the mechanical bath eventually destroys coherent correlations between mechanics and light. We therefore opt for pulses with an effective duration that is smaller than the thermal decoherence time $\Gamma_\rmmath{th}^{-1}$, cf. \eref{eq_Gammath}. As a variational class of mode functions we choose exponentially decaying envelopes for the demodulating phases, see \fref{fig_modeFunct},
\begin{subequations}\label{eq_modeFunct_time}
\begin{align}
    f_\rmmath{E}(t) &\coloneqq N e^{(\Gamma -\i \omega_\rmmath{m}) t} \theta\left(-t- T_\rmmath{sep}/2 \right),\\
    f_\rmmath{L}(t) &\coloneqq N e^{(-\Gamma +\i \omega_\rmmath{m}) t} \theta(t -T_\rmmath{sep}/2),
\end{align}
\end{subequations}
where the decay rate $\Gamma$ scales the effective duration of the mode, $\theta$ is the Heaviside step function and $N=(2\Gamma e^{\Gamma T_\rmmath{sep}})^{1/2}$ is the normalization constant. This class of mode functions is in particular motivated by Ref.~\cite{Miao1}, where it is shown that the temporal mode of the light sharing maximal amount of entanglement (largest negativities) with the oscillator have an exponentially decaying envelope for a demodulating phase at the mechanical frequency $\omega_\rmmath{m}$. These mode functions are anti-causal and causal, respectively, and they are orthonormal by construction. The pulse decay rate $\Gamma$ is a free variational parameter with respect to which we will minimize $\EPR$.

\subsection{Experimental task}\label{sec_ExpTask}
\sref{sec_entVerif} defines a readily implementable scheme to detect optomechanical entanglement between temporally ordered modes of the output field. The optomechanical cavity is driven on resonance in the non-sideband-resolved regime by a constant input laser while the cavity output field is continuously monitored. The choice of the exact measurement scheme depends on the entanglement criterion of interest. Here \eref{eq_EPRvar_quad} [or equivalently \eref{eq_EPRvar_ladderOp}] prescribes what information the detection scheme in an experiment must provide. For example, the equal-time correlators in \eref{eq_EPRvar_quad} can be evaluated from records of two homodyne measurements of the phase and the amplitude of the output field. From such a record, temporally ordered modes can be extracted according to \eref{eq_modeOp_time}. If the dynamics are stationary, the detection of entanglement between the light modes implies that the mechanical oscillator and the output light share stationary entanglement. \Fref{fig_scheme} summarizes this procedure.

\section{RESULTS}\label{sec_results}
This section has two parts: first, we provide an approximate formula for the EPR-variance and comment on what it teaches us about entanglement in the system; second, we benchmark the approximate formula with its exact counterpart. The latter is very cumbersome in its symbolic form and, on its own, provides only numerical results, whereas our approximate formula gives generic behavior with respect to the parameters.

\subsection{Approximate formula for the EPR-variance}
Using the mode functions of \esref{eq_modeFunct_time} in the definition of the mode operators in \eref{eq_modeOp_time}, the EPR-variance in \eref{eq_EPRvar_ladderOp} can be evaluated by means of the input--output relations \esref{eq_outputQuad} and the noise correlators \esref{eq_mechBath_correl} and \eqref{eq_lightBath_correl}. A detailed derivation is given in \aref{sec_DerivEPRvarGeneral}.

Here we give the final result, which is compared to (and backed up by) exact numerical calculations in \sref{sec_compar_ex_num},
\begin{widetext}
\begin{equation}\label{eq_EPRformula}
    \EPR = 2 + \eta \frac{4 \GammaRO}{\Gamma + \frac{\gamma_\mathrm{m}}{2}}\left[ \frac{2\left({\GammaRO} + \Gamma_\mathrm{th} \right)}{\gamma_\mathrm{m}} \left(1 - \frac{\Gamma e^{-\gamma_\rmmath{m}T_\rmmath{sep}/2}}{\Gamma + \frac{\gamma_\mathrm{m}}{2}} \,\cos \phi \right) -  \left(\frac{\Gamma e^{-\gamma_\rmmath{m}T_\rmmath{sep}/2}}{\Gamma + \frac{\gamma_\mathrm{m}}{2}} \,\cos \phi \right) \right].
\end{equation}
\end{widetext}
The readout and thermal decoherence rate, $\GammaRO$ and $\Gamma_\text{th}$, have been defined in \esref{eq_GammaRO} and \eqref{eq_Gammath}. We included a finite detection efficiency $\eta \leq 1$. Equation \eqref{eq_EPRformula} is derived for a resonant drive $\delta =0$ and the regime $\kappa \gg \omega_\rmmath{m} \gg \gamma_\rmmath{m}$ such that approximations \esref{eq_approximatedFunctions} apply. Furthermore, in order to arrive at an analytically tractable formula, we made the technical assumption that the pulse envelopes decay at a rate $\Gamma$ fulfilling
\begin{equation}\label{eq_approxGamma}
    \Gamma \ll\frac{\omega_\rmmath{m}}{\sqrt{n_\rmmath{th}(\Cq+1)}}\approx\frac{\omega_\rmmath{m}}{\sqrt{\mathcal{C}_\rmmath{cl}+n_\rmmath{th}}}.
\end{equation}
Here $\Cq$ is the optomechanical quantum cooperativity, cf. \eref{eq_cooperativity}, and $\mathcal{C}_\rmmath{cl} \coloneqq 4g^2/\kappa\gamma_\mathrm{m}$ is the classical cooperativity. The approximation assumes the large temperature limit. Equation \eqref{eq_approxGamma} has to be read as a limitation on the optomechanical coupling $g$. As we will see, for couplings that violate \eref{eq_approxGamma} the EPR-variance is not a suitable entanglement criterion anymore. Thus, in the regime in which the EPR-variance is of interest, \eref{eq_approxGamma} is a self-consistent restriction.

It is useful to characterize the circumstances for which $\EPR$ is minimal, so that the violation of the separability bound, \eref{eq_DuanCrit}, is least susceptible to inevitable experimental imperfections. To this end, we set $\phi=0$ in \eref{eq_EPRformula}. The first term in the square brackets is always positive and monotonically decreasing with the decay rate of the temporal modes $\Gamma$. The second term is negative and monotonically increasing with $\Gamma$. Only if the second term can compensate the first one is it possible to detect entanglement. In view of this and the overall scaling in front of the square brackets with respect to $\Gamma$, it is clear that there is an optimal pulse bandwidth yielding an optimal EPR-variance.

We discuss first the result of this minimization for ideal detection, $\eta=1$, and vanishing pulse separation, that is, in zeroth order of $\gamma_\rmmath{m} T_\rmmath{sep}$.  A straightforward calculation gives the optimal pulse bandwidth
\begin{equation}\label{eq_OptGamma}
\begin{split}
    \Gamma_\rmmath{opt} &= 2({\GammaRO} + \Gamma_\rmmath{th}) +\frac{\gamma_\rmmath{m}}{2}\\ &\approx\  2 \gamma_\rmmath{m}(\mathcal{C}_\mathrm{cl}+n_\mathrm{th}),
\end{split}
\end{equation}
for which the approximation holds when $n_\mathrm{th}\gg 1$. We expected that the pulses have to decay faster than the thermal decoherence rate in order to preserve the coherence between the pulses with large probability. This means that the pulses' bandwidth can not be too narrow. The factor of 2 accounts for the fact that the coherence must be kept  across both pulses, i.e., there may be no significant incoherent leakage over times $\sim 2/\Gamma_\rmmath{opt}$. Increasing the readout rate (or the cooperativity) leads to shorter optimal pulses. We interpret this tendency as a tentative to extract information from the output light as fast as the dynamics allows, in order to minimize the chances of decoherence. This seems a natural strategy because, in our model, thermal decoherence is the only channel through which entanglement can be lost. However, this might not be the best strategy in a real experiment: as pulses become shorter, their bandwidth  broadens and incoherent spectral features from technical noise and/or additional mechanical modes will be resolved by the pulses. We expect that this will prevent observation of entanglement if not accounted and controlled carefully \cite{Thesis_Jason}. Moreover, the form of the optimal bandwidth $\Gamma_\rmmath{opt}$ is only valid when \eref{eq_approxGamma} holds, which limits the maximal bandwidths we can predict this way. The optimal pulse bandwidth in \eref{eq_OptGamma} is compatible with the assumption in \eref{eq_approxGamma}, if $2(\mathcal{C}_\mathrm{cl}+n_\mathrm{th})^{3/2}\ll Q$. For large optomechanical coupling $g$, which violate this condition, entanglement will not be detectable in the form of EPR-squeezing and one should instead consider more general entanglement witnesses, as we will see.

Formula \eqref{eq_EPRformula}, for the optimal choice of bandwidth $\Gamma_\rmmath{opt}$, predicts
\begin{equation}\label{eq_EPRformulaOpt}
    \EPR = 1+ \frac{1}{\Cq+1}.
\end{equation}
 For arbitrarily small (but finite) cooperativities, this expression is always smaller than two. This means that entanglement between the light modes is always present and so is optomechanical entanglement. This finding is consistent with the results of Ref.~\cite{Miao1}, where this surprising fact has been noted first. Equation \eqref{eq_EPRformulaOpt} constitutes the main result of our work. For large quantum cooperativity $\EPR$ approaches $50\%$ of squeezing below shot noise. This limit hints at the fact that we consider only two (out of infinitely many) light modes that encode correlations with the mechanics.

When finite detection efficiency and pulse separation are taken into account, the minimal EPR-variance, \eref{eq_EPRformulaOpt}, is modified as follows (to first order in $\gamma_\rmmath{m}T_\rmmath{sep}$)
\begin{equation}\label{eq_EPRformulaOpt1}
    \EPR = 2- \eta \frac{\Cq}{\Cq+1}+4\eta \GammaRO T_\rmmath{sep}.
\end{equation}
In the relevant regime of large cooperativity, $\Cq>1$, we have $\GammaRO>\Gamma_\rmmath{th}$ such that the pulse separation $T_\rmmath{sep}$ needs to stay well below the thermal decoherence time $\Gamma_\rmmath{th}^{-1}$, as one would expect.

From the derivation detailed in \aref{sec_DerivEPRvarGeneral}, it is possible to give a detailed account on the physical origin of each term in \eref{eq_EPRformula}. The factor 2 at the front is the shot noise contribution of the light field that we would observe in the absence of optomechanical coupling (i.e., if ${\GammaRO}=0$). The first term in the square brackets is due to auto-correlations in thermal noise and in back-action noise, contributing to the two pulses. This term has two parts: a non-negative part coming from intra-pulse correlation in each of the two temporal modes (early or late with themselves), and a negative part coming from the inter-pulse correlation between early and late mode. The net effect of both parts is always positive, therefore auto-correlations in thermal noise and back-action noise do not contribute to entanglement. The last term in the square brackets corresponds to cross-correlations of shot noise and back-action noise. These correlations are also at the basis of ponderomotive squeezing typically observed in the frequency domain \cite{Safavi-Naeini_2013,Purdy2013a}. Here, the correlations refer purely to inter-pulse correlation of the two temporal modes; the intra-pulse correlation gives an exact zero contribution, as detailed in \aref{sec_DerivEPRvarGeneral}. In this sense, an EPR-variance below 2 is due to ponderomotive squeezing between the early and late modes.

\subsection{Comparison with numerical results}\label{sec_compar_ex_num}

Ref.~\cite{Genes_2008} provides a procedure to express analytically (in integral form) the covariance matrix of arbitrary temporal modes of the continuous steady-state output field of an optomechanical cavity. Arbitrary parameter regimes (e.g., nonzero detuning,  sideband resolving cavity, strong coupling, etc.) can be studied numerically this way, as long as the system is stable and reaches a steady state. The form of the mode functions defined in \esref{eq_modeFunct_time}, for vanishing $T_\rmmath{sep}$, allows one to compute symbolically the \emph{exact} expression of the covariance matrix, and the EPR-variance is readily obtained from the entries of the latter. The exact (symbolic) formula for the EPR-variance is cumbersome and not very informative in its symbolic form, but it provides (exact) numerical results upon evaluation with fixed parameters; in this section numerical results refer to the evaluation of the exact symbolic expression with numerical values. We compare below the approximate formula, \eref{eq_EPRformula}, to this approximation-free method.

In the following, unless otherwise stated, we assume $\omega_\rmmath{m}/2\pi= 1\ \mathrm{MHz}$, $\kappa=10\,\omega_\rmmath{m}$, $\mathrm{Q}=\omega_\rmmath{m}/\gamma_\rmmath{m}=10^8$, and $n_\rmmath{th}=10^4$. This parameter set is partly inspired by a recent experiment \cite{rossi2018observing}. All plots refer to the case $\eta=1$ and $T_\rmmath{sep}=0$ (which corresponds to the limit where $T_\rmmath{sep}$ does not significantly alter $\EPR$).  The optimal angle $\phi_{\rmmath{opt}}$ for the exact EPR-variance is found analytically and used throughout all figures. The scripts and data for generating the plots are freely available online \cite{gut_corentin_2019_3901001}.

\begin{figure}
\includegraphics[width=\columnwidth]{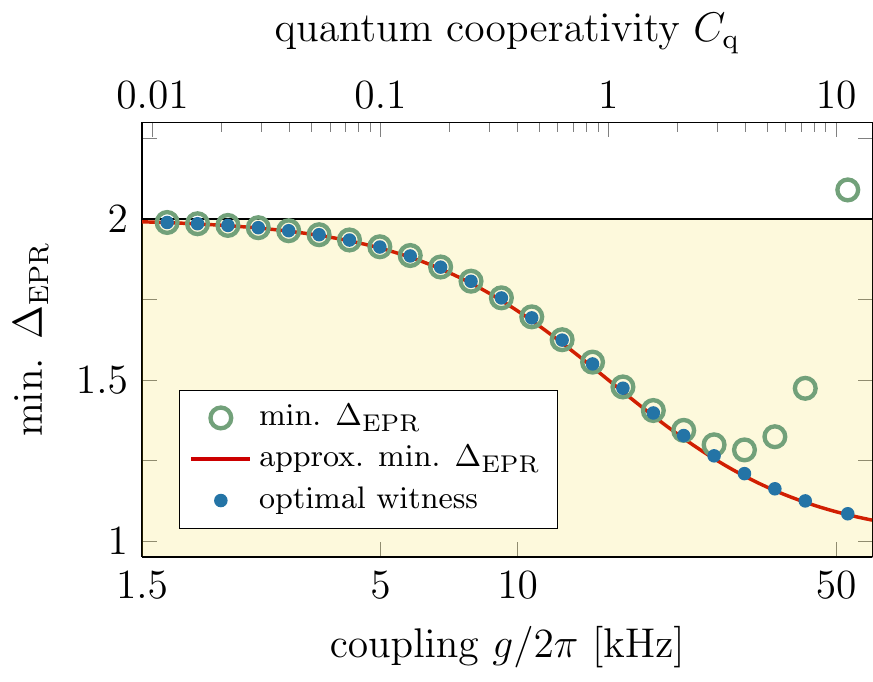}
\caption{Approximate minimal EPR-variance from \eref{eq_EPRformulaOpt} (red solid line), exact EPR-variance (green circles) and optimal entanglement witness (blue dots) versus optomechanical coupling $g$ (lower $x$-axis) and quantum cooperativity $\Cq$ (upper $x$-axis). Threshold for separable states is at 2 (solid black line), the yellow shaded area represents the region of entanglement for the EPR-variance and the optimal witness. This applies to all other figures.} \label{fig_minEPR_G}
\end{figure}

\Fref{fig_minEPR_G} shows the minimal EPR-variance versus optomechanical coupling $g$ (and quantum cooperativity $\Cq$), comparing the approximate formula, \eref{eq_EPRformulaOpt} (red line), to the exact results (green circles) of the approximation-free symbolic expression. In the latter, for each $g$, we sweep over $\Gamma$ to determine the minimal value of $\EPR$. For cooperativities $\Cq \lesssim 3$ the plots overlap well. In particular, the exact result confirms the surprising presence of entanglement at small cooperativities.  The approximate formula becomes inaccurate at larger cooperativities because the restriction on the pulse bandwidths, \eref{eq_approxGamma} for $\Gamma=\Gamma_\rmmath{opt}$, does not hold any more: in \fref{fig_minEPR_G}, the approximate $\EPR$ predicted by the red curve departs from the exact results given by the green circles. The exact $\EPR$ rises above the separability threshold 2 when the cooperativity increases  because the EPR-variance criterion is not necessary for entanglement and fails to detect that certain states are entangled. For every two-mode Gaussian state (the class of states the temporal modes of the light belong to, recall \sref{sec_entCriterion}) it is possible to find the \textit{ optimal} entanglement witness, based on linear combinations of second moments (covariances), that decides whether the state is entangled or not  \cite{Hyllus_2006}\footnote{Ref.~\cite{Hyllus_2006} provides a \textsc{MATLAB} function that computes the optimal witness given a covariance matrix.}. Because the EPR-variance is a (in general sub-optimal) witness based on second moments  as well, we can plot the optimal witness on the same $y$-axis with the same separability thresholds at 2. The blue dots in \fref{fig_minEPR_G} are the optimal-witness values  of the states computed exactly. They monotonically decrease as $\Cq$ increases,  therefore confirming the expected behavior that larger coherent coupling does not worsen the detection of entanglement. Interestingly, the approximate formula for the EPR-variance and the optimal witness overlap well. We interpret this to be a result of the general scaling of squeezing as $1/\Cq$ in the limit of large cooperativity.

\begin{figure}
\includegraphics[width=\columnwidth]{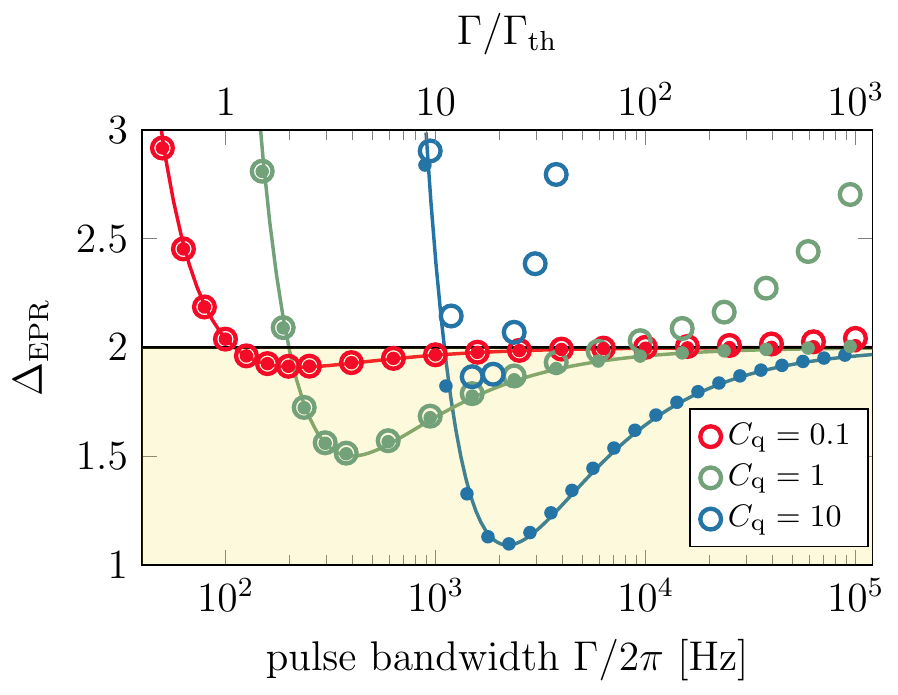}
\caption{Approximate minimal EPR-variance from \eref{eq_EPRformulaOpt} (solid lines), exact EPR variance (circles), optimal entanglement witness (dots) versus pulse bandwidth $\Gamma$ for cooperativities of $\Cq= 0.1,\,1,\,10$ in red, green, blue, respectively.} \label{fig_EPR-Gamma}
\end{figure}

\Fref{fig_EPR-Gamma} shows how the EPR-variance changes with the pulse bandwidths $\Gamma$ according to the approximate formula, \eref{eq_EPRformula} (solid lines), and compared to the exact results (circles). All curves display a single global minimum that is not a sharp feature. Thus the optimisation over the pulse bandwidths is not too difficult in principle; especially in an experimental scenario where all the parameters are not known exactly. According to formula \eqref{eq_EPRformula}, larger pulse bandwidths will always yield some entanglement though one has to keep in mind the restriction of \eref{eq_approxGamma} that limits the validity of the formula with respect to large $\Gamma$. As mentioned in the previous section, thermal noise on the mechanics is the sole decoherence channel of our model, therefore pulses decaying faster than the thermal decoherence time $1/\Gamma_\rmmath{th}$ will display some coherence. In practice, broadband pulses (short in time) will resolve other incoherent spectral components (electronic filters, additional mechanical modes, etc.) not accounted for in the present model. Close to the minima, we see good agreement for $\Cq \lesssim 1$, consistently with \fref{fig_minEPR_G}. The exact EPR-variance attains larger values at larger cooperativity because it is a sub-optimal witness, as in \fref{fig_minEPR_G}. On the curve $\Cq=1$, the approximate formula and the exact results agree across the minimum and no more at larger bandwidths where the restriction of \eref{eq_approxGamma} does not hold.

\begin{figure}
\includegraphics[width=\columnwidth]{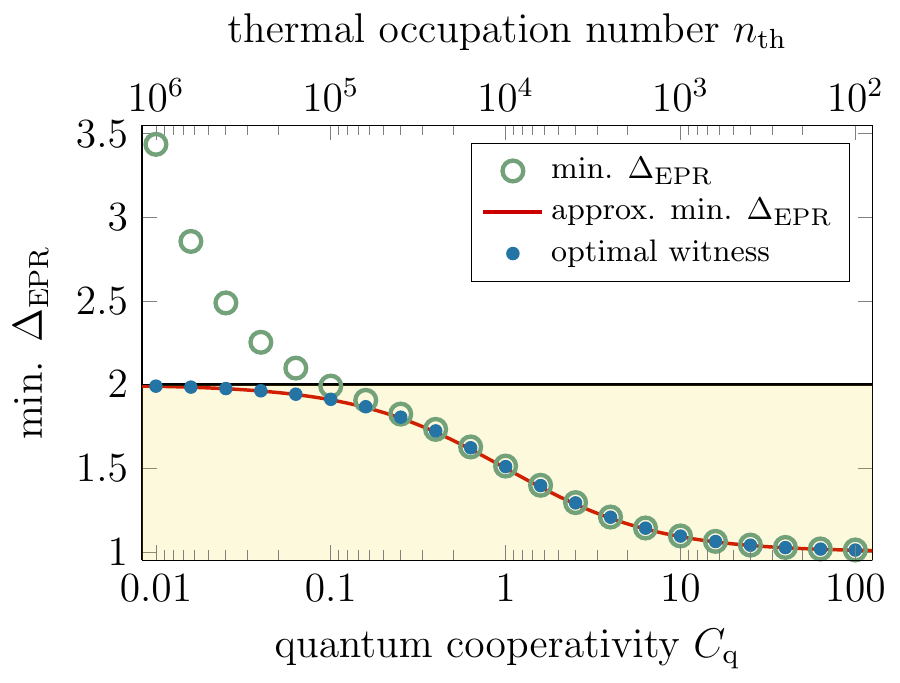}
\caption{Approximate minimal EPR-variance from \eref{eq_EPRformulaOpt} (red solid line), exact EPR variance (green circles), optimal entanglement witness (blue dots) versus mechanical bath temperature, parametrized in terms of the mean thermal occupation number, for fixed optomechanical coupling $g/2\pi=15.8\rmmath{kHz}$}.
\label{fig_minEPR_nth}
\end{figure}

In \fref{fig_minEPR_nth} we illustrate the dependence of the minimal EPR-variance and the optimal witness on the temperature of the mechanical oscillator's  bath, parametrized by its mean thermal occupation $n_\mathrm{th}$. The cooperativity increases like the inverse of $n_\mathrm{th}$. The approximate formula, \eref{eq_EPRformulaOpt} (red line), and the exact numerical result (green circles) agree when $\Cq>1$. The exact $\EPR$ and the optimized witness appear to saturate at a minimal value of $50\%$ of squeezing below shot noise, just like the approximate formula predicts. Formula \eqref{eq_EPRformula} becomes inaccurate as $n_\mathrm{th}$ grows large where, again, the restriction of \eref{eq_approxGamma} does not hold any more. At large temperature of the mechanical bath, the optimal witness (blue dots) predicts retrievable entanglement, which is similar to the behavior at small coupling on \fref{fig_minEPR_G}.

\begin{figure}
\includegraphics[width=\columnwidth]{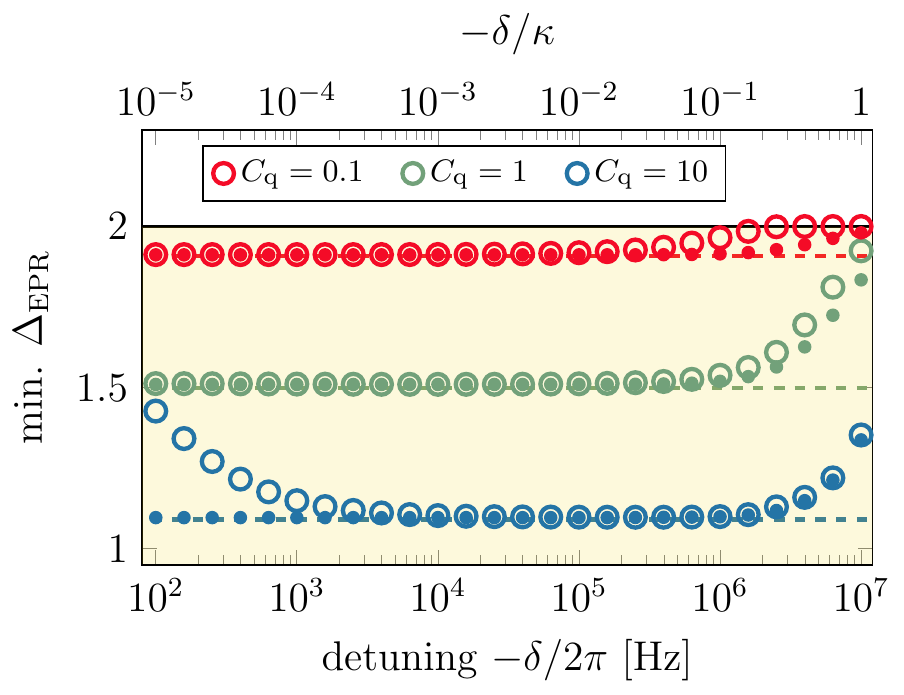}
\caption{Exact EPR variance (circles) and optimal entanglement witness (dots) versus red detuning of the driving field from cavity resonance. The optomechanical coupling is kept constant corresponding to a quantum cooperativity of $\Cq=0.1,\,1,\,10$ for red, green, and blue symbols, respectively. The dashed lines are the approximate minimal EPR-variance formula, \eref{eq_EPRformulaOpt}, valid only at $\delta =0$ (displayed for comparision).}\label{fig_minEPR_delta}
\end{figure}

In realistic experiments, zero detuning is a challenging regime of operation because it is at the border to the unstable regime of the optomechanical dynamics at blue detunings. Therefore, in practice, the laser drive is usually slightly red detuned from the cavity resonance. In \fref{fig_minEPR_delta} we study the effects of detuning on the minimal EPR-variance with the exact expressions of the covariance matrix: the circles correspond to the exact $\EPR$ and the dots to the optimal-witness values. For comparison, the values of formula \eref{eq_EPRformulaOpt} (only valid for zero detuning) are displayed as  dashed lines. As the detuning approaches the cavity linewidth $\kappa$ the separability bound violation diminishes because the strength of the sidebands' spectral components in the cavity output field is reduced by the cavity profile. This is analogous to decreasing the detection efficiency (passive losses), in the sense that the spectral weight of the signal extracted by the mode functions decreases relative to the shot noise component. On the other hand, detuning does not introduce any additional noise, therefore the EPR-variance does not increase above 2. In \fsref{fig_minEPR_G} and \ref{fig_EPR-Gamma}, where the detuning was zero, the exact EPR-variance increases with cooperativity and \fref{fig_minEPR_delta} shows this too for large cooperativities and small detunings. Remarkably, at large cooperativity, some detuning reduces the minimal EPR-variance down to an asymptotic value close to the prediction of the approximate formula \eref{eq_EPRformulaOpt} for zero detuning.

\subsection{Remarks on shot-noise levels and multi-mode entanglement}\label{sec_shotnoiselevel}

In this section, we comment on the -- somewhat subtle --
role of shot noise levels in experimental entanglement
detection in optomechanical systems.
Given the canonical coordinates $O=(x_\mathrm{L},p_\mathrm{L},
x_\mathrm{E},p_\mathrm{E})$, one can capture the
second moments in
a $4\times 4$ covariance matrix $\Xi$
with entries
\begin{equation}
    \Xi_{i,j} \coloneqq \langle O_i O_j + O_j O_i\rangle,
\end{equation}
for vanishing means or first moments
$\langle O_j\rangle$ for $j=1,\dots, 4$.
Any covariance matrix of a quantum state satisfies
the Heisenberg uncertainty principle
\cite{Continuous,Adesso_2007,GaussianQuantumInfo}
\begin{equation}
    \Xi+\i\sigma\geq 0,
\end{equation}
where $\sigma$ is the symplectic matrix incorporating the canonical commutation relations.

We will now turn to entanglement criteria.
Any entanglement
test linear in $\Xi$ can be written in the
form
\begin{equation}\label{Form}
    {\rm tr}(X \Xi)<1,
\end{equation}
where $X$ is the $4\times 4$ matrix that
captures the measurement settings. It takes a
moment of thought -- and is explained in
Ref.~\cite{Hyllus_2006} -- that the Duan
criterion
\begin{equation}
\Delta_{\rm EPR}={\rm tr}(X \Xi)
\end{equation}
of Ref.~\cite{Duan_2000} corresponds
to a specific such choice for $X$. Such tests are not only convenient for their simplicity in experimental implementations: their linear nature in $\Xi$ renders the assessment of confidence regions of tests simple and feasible.

There is a subtle aspect that one needs to keep in mind when experimentally verifying entanglement: for a bi-partite Gaussian state, a state being entangled and having a positive partial transpose is one-to-one. That is to say,
exactly the covariance matrices of separable
Gaussian quantum states will correspond to
covariance matrices that satisfy
\begin{equation}\label{CM}
    \Xi^\Gamma +\i\sigma\geq 0,
\end{equation}
where $\Xi^\Gamma$ is the covariance matrix of the partial transpose. A slightly inaccurate
experimental
assessment of the shot noise level will
correspond to a matrix $x \sigma$ where the real $x$ is slightly different from unity. That means that when in such a picture the matrix $\Xi$ is slightly unphysical,
\begin{equation}
\Xi+\i x \sigma\not\geq 0,
\end{equation}
it seemingly comes along with the Gaussian state being certified as being entangled
\begin{equation}
    \Xi^\Gamma +\i x\sigma\not \geq 0,
\end{equation}
even if it is not. Therefore, to certify entanglement with confidence it is crucial to obtain a precise shot noise characterisation, taking into account statistical fluctuations and any temporal drift. Note that these effects can be highly relevant even if the states involved are close to pure. In yet other words, if covariance matrices recovered are close to being pure [and hence close to the boundary of the convex set of covariance matrices defined by \eref{CM}], enormous care is necessary when drawing the conclusion that the state is entangled: this may be an artefact of an insufficiently calibrated shot noise level.

This issue is aggravated in the multi-mode case, which is specifically interesting when assessing entanglement between several optical and mechanical modes. Entanglement tests still take the form of \eref{Form}, now $\Xi$ being a symmetric $2n\times 2n$ matrix for $n$ modes in total \cite{Hyllus_2006}. Such tests are highly convenient for detecting multi-mode entanglement in optomechanical systems, in that the optimal test for an anticipated quantum state can be efficiently found by means of methods of semi-definite programming \cite{Hyllus_2006}, so
convex optimization techniques \cite{Boyd2004}.

The above
mentioned situation is now expected to be generic: in common experiments, some modes of local subsystems will be close to being pure. A
more sophisticated statement of this kind is that in similar ways as natural quantum states are close to being low-rank states \cite{Compressed},
commonly encountered covariance matrices $\Xi$
will have a low quantity $r$ that can be seen as a ``symplectic rank'': if
$m$ is the number of unit eigenvalues of
$-(\Xi\sigma)^2$, then
\begin{equation}
    r=n-m.
\end{equation}
A pure Gaussian state will have $r=0$, a state each mode of which is mixed $r=n$. In this language,
commonly encountered Gaussian states will feature covariance matrices that are
well approximated by covariance matrices that
have a value of $r$ different from $n$: so some modes will be very susceptible to inaccurately estimated shot noise levels. If they are
slightly underestimated, then most naturally
encountered states will seem entangled, even if there is no entanglement in the system. This is an important observation little appreciated so far in the literature. This observation also comes hand in hand with the fact that in practical recoveries, a slight unphysicality of the covariance matrices goes hand in hand with entanglement being detected. Having said that,
this applies only to the situation of inaccurately assessed shot
noise levels: in some setting, even a weak signal in optomechanical systems subjected to very high
noise levels can lead to strongly significant predictions \cite{Observation}.
Again, this is an aspect one has to be very
careful about when assessing optomechanical
entanglement.

\section{CONCLUSION}
In this work, we have presented an argument to demonstrate the presence of stationary optomechanical entanglement in a continuously driven optomechanical cavity. The argument relies on the extraction of temporally ordered modes from the continuous photocurrent of homodyne detection in post-processing -- no modulation of the drive is needed. The analytic study of a specific scheme shows that EPR-squeezing can reach up to 50\% below the threshold of separable states for large quantum cooperativity $\Cq$. Our approximate formula for the EPR-variance predicts that this limit is approached as $\Cq^{-1}$. Remarkably, this limit seems to hold also for the exact EPR-variance as well as for optimal entanglement witnesses which are linear in the second moments. It would be interesting to establish this as a strict limit.

We have studied a specific class of variational mode profiles, which are physically well motivated, and are theoretically capable to reveal entanglement. Nevertheless, a systematic optimization of the mode functions would be desirable. This could be based, e.g., on the approach indicated in the end of \aref{sec_exactEPRformula}.

The single-mode model of the mechanical oscillator has been sufficient to fully develop our entanglement verification argument. Now, single-mode resonators are rare and it turns out that our scheme can be generalized to multiple mechanical modes: for each modes an early and late pair of mode operators, \eref{eq_modeOp_time}, are defined such that the covariance matrix, of dimension twice the number of modes included, can be computed. Entanglement in this state can be assessed with respect to the bi-partition early--late. The inclusion of more modes improves the purity of the reconstructed state, in the sense that spectral contributions from the included modes are treated as signal rather than strong noise. This procedure (original and detailed study in Ref.~\cite{Thesis_Jason})  might significantly improve experimental attempts, together with the considerations on physical states reconstruction of \sref{sec_shotnoiselevel}.

We hope that our scheme provides a viable pathway to demonstrate stationary optomechanical entanglement, as predicted over a decade ago \cite{Genes_2008}.  It would open a way for sophisticated protocols providing a full characterization of the optomechanical entanglement from, e.g., state tomography. We think that a successful implementation of our scheme would constitute a textbook experiment exploring the Heisenberg--von Neumann cut between measured system and measurement apparatus in the example of an optomechanical position meter.

\section*{ACKNOWLEDGMENTS}
The authors are very thankful to Claus G{\"a}rtner and Marek Gluza for important discussions and inputs in the course of this project.
C.~G.\ acknowledges support from the European Union’s Horizon 2020 research and innovation program under the Marie Sklodowska-Curie grant agreement No.\ 722923 (OMT).
N.~W.\ acknowledges funding support from the European Unions Horizon 2020 research and innovation programme under the Marie Sklodowska-Curie grant agreement No.\ 750905.
J.~E.\ acknowledges support by the DFG (CRC183, FOR 2724), and the FQXi.
M.~A.\ acknowledges support by the European Research Council (ERC QLev4G). This project was supported by the doctoral school CoQuS (Project W1210).
K.~H.\ acknowledges support through DFG through CRC 1227 DQ-mat Project No. A06 and Germany’s Excellence Strategy – EXC2123 QuantumFrontiers – 390837967

\appendix

\section{APPENDIX: LIGHT-LIGHT ENTANGLEMENT IMPLIES LIGHT-MECHANICS ENTANGLEMENT} \label{sec_ArgumentDetails}
In this Appendix we provide the details of the indirect proof of entanglement in the scenario shown in \fref{fig_quantumCircuit}. The initial state is a product state of E, M, and L,
\begin{align}
  \rho^\mathrm{init}_\mathrm{EML}= \rho_\mathrm{E}\otimes\rho_\mathrm{M}\otimes\rho_\mathrm{L}.
\end{align}
The first quantum channel $\mathcal{E}_\mathrm{EM}$ acts non-trivially on E and M only. If this quantum channel does \emph{not} generate entanglement between E and M, then
\begin{align}
  (\mathcal{E}_\mathrm{EM}\otimes\mathds{1}_\mathrm{L})\rho^\mathrm{init}_\mathrm{EML}&
  =\mathcal{E}_\mathrm{EM}(\rho_\mathrm{E}\otimes\rho_\mathrm{M})\otimes\rho_\mathrm{L} \nonumber\\
  &=\sum_{i} p_i \rho_\text{E}^{i}\otimes\rho_\mathrm{M}^{i}\otimes\rho_\mathrm{L}. \label{eq_E_EM}
\end{align}
In this case, any subsequent channel acting on M and L cannot generate entanglement between E and L: the final state is
\begin{align}
  \rho_\mathrm{EML}^\mathrm{fin}&=(\mathds{1}_\mathrm{L}\otimes\mathcal{E}_\mathrm{ML})
  (\mathcal{E}_\mathrm{EM}\otimes\mathds{1}_\mathrm{L})\rho^\mathrm{init}_\mathrm{EML} \nonumber\\
  &=\sum_{i} p_i \rho_\text{E}^{i}\otimes\mathcal{E}_\mathrm{ML}(\rho_\mathrm{M}^{i}\otimes\rho_\mathrm{L})
\end{align}
and the reduced state of E and L is
\begin{align}
  \rho_\mathrm{EL}^\mathrm{fin}&= \mathrm{tr}_\mathrm{M}\left[\rho_\mathrm{EML}^\mathrm{fin}\right]
  =\sum_{i} p_i  \rho_\mathrm{E}^{i}\otimes\tilde{\rho}_\mathrm{L}^{i}
\end{align}
where $\tilde{\rho}_\mathrm{L}^{i}=\mathrm{tr}_\mathrm{M}\left[\mathcal{E}_\mathrm{ML}(\rho_\mathrm{M}^{i}\otimes\rho_\mathrm{L})\right]$. The state $\rho_\mathrm{EL}^\mathrm{fin}$ is separable. In essence, this statement is equivalent to the obvious fact that entanglement cannot be generated by local operations. In section \ref{sec_argument} we use the contraposed statement: if $\rho_\rmmath{EL}^\rmmath{fin}$ is entangled, then $\mathcal{E}_\mathrm{EM}$ must have created E--M entanglement.

 In the literature we find that, with the same pairwise interaction sequence, it is possible to distribute entanglement between parties E and L via a mediator M, in such a way that M is never entangled with E and/or L \cite{Cubitt_2003, Korolkova_08, Fedrizzi_2013, Peuntinger_2013, Vollmer_2013}. There, the initial state of E, M and L is a separable state but it crucially \textit{is} classically correlated. It is because we require that the initial state is uncorrelated that we can write in general that the action of $\mathcal{E}_\mathrm{EM}$ yields a separable state in \eqref{eq_E_EM}.

\section{APPENDIX: DERIVATION OF ANALYTICAL FORMULA FOR THE EPR-VARIANCE}\label{sec_DerivEPRvarGeneral}
This section presents the derivation of formula \eref{eq_EPRformula} for the EPR-variance for the exponentially decaying mode functions defined in \esref{eq_modeFunct_time}, valid in the regime  $\kappa \gg \omega_\rmmath{m} \gg \gamma_\rmmath{m}$, $\Gamma \ll \omega_\rmmath{m}/\sqrt{n_\rmmath{th}(C+1)}$ and $\delta =0$. The methodology below is strongly inspired from \cite{Genes_2008}.

\subsection{Solution of the Langevin equations in Fourier space}\label{sec_SolQLE}

In this document, the definition of the Fourier transform of a function or operator $h$ is
\begin{subequations}
\begin{align}\label{eq_ConventionFT}
\mathcal{F}[h(t)]_\omega& \coloneqq \int_{-\infty}^{\infty} \frac{\rmmath{d}t}{\sqrt{2\pi}} e^{\i\omega t} h(t) = h(\omega),\\
\mathcal{F}^{-1}[h(\omega)]_t& \coloneqq \int_{-\infty}^{\infty} \frac{\rmmath{d}\omega}{\sqrt{2\pi}} e^{-\i\omega t} h(\omega) = h(t).
\end{align}
\end{subequations}
In the operator context we define that the the adjunction is performed after the Fourier transformation: we write
\begin{equation}
    h^\dagger(\omega) \coloneqq \left( h(\omega) \right)^\dagger \coloneqq \left(\mathcal{F}[h]_\omega \right)^\dagger = \mathcal{F}\left[h^\dagger \right]_{-\omega}.
\end{equation}

The Fourier transform of the QLE, \esref{eq_QLEtime}, with $\delta = 0$, are
\begin{subequations}\label{eq_QLEfreq}
\begin{align}
    -\i \omega x_\rmmath{m}(\omega)&= \omega_\rmmath{m} p_\rmmath{m}(\omega)\label{eq_QLEfreq1},\\
    -\i \omega p_\rmmath{m}(\omega)&= -\gamma_\rmmath{m} p_\rmmath{m}(\omega) -\omega_\rmmath{m} x_\rmmath{m}(\omega) \nonumber\\ &\quad -2gx_\rmmath{c}(\omega) +\sqrt{2\gamma_\rmmath{m}} \xi(\omega) \label{eq_QLEfreq2}, \\
    -\i \omega x_\rmmath{c}(\omega)&= -\frac{\kappa}{2}x_\rmmath{c}(\omega) + \sqrt{\kappa}x_\rmmath{in}(\omega) \label{eq_QLEfreq3},\\
    -\i \omega p_\rmmath{c}(\omega)&= -\frac{\kappa}{2} p_\rmmath{c}(\omega) -2gx_\rmmath{m}(\omega) \nonumber\\
    &\quad + \sqrt{\kappa}p_\rmmath{in}(\omega)\label{eq_QLEfreq4}.
\end{align}
\end{subequations}
We express $x_\rmmath{c}$ and $p_\rmmath{c}$ in terms of the driving noises: electromagnetic vacuum fluctuations (shot noise) $x_\rmmath{in}, p_\rmmath{in}$ and Hermitian thermal mechanical noise $\xi$. With the input--output relations, $x_\rmmath{out} = \sqrt{\kappa} x_\rmmath{c} - x_\rmmath{in}$ (analogously for $p_\rmmath{out}$) \cite{Gardiner_Collet_85}, we find the quadratures of the light escaping the cavity, \esref{eq_outputQuad},
\begin{align*}
    x_\rmmath{out}(\omega) &= \mathcal{S}(\omega) x_\rmmath{in}(\omega),\\
    p_\rmmath{out}(\omega) &= \mathcal{S}(\omega) p_\rmmath{in}(\omega) + 4g^2 \chi_\rmmath{opt}^2(\omega) \chi_\rmmath{m}(\omega) x_\rmmath{in}(\omega) \nonumber\\
    &\quad- 2g \chi_\rmmath{opt}(\omega) \chi_\rmmath{m}(\omega) \sqrt{2\gamma_\rmmath{m}} \xi (\omega).
\end{align*}

It will be convenient to use the following abbreviations for the pre-factors of shot noise, back-action noise, and thermal noise, respectively,
\begin{subequations} \label{eq_noiseFactors}
    \begin{align}
        \mathcal{S}(\omega)&\coloneqq \frac{\frac{\kappa}{2} +\i \omega}{\frac{\kappa}{2} -\i \omega},\\
        \mathcal{B} (\omega)&\coloneqq 2g^2 \chi_\rmmath{opt}^2(\omega) \chi_\rmmath{m}(\omega),\\
    \mathcal{T}(\omega) &\coloneqq 2g \sqrt{\gamma_\rmmath{m}(n_\mathrm{th} + 1/2)} \chi_\rmmath{opt}(\omega) \chi_\rmmath{m}(\omega).
    \end{align}
\end{subequations}
These functions obey $\mathcal{S}^*(\omega) = \mathcal{S}(-\omega)$, and similarly for $\mathcal{B}(\omega)$ and $\mathcal{T}(\omega)$. The optical and mechanical susceptibilities were introduced in \esref{eq_CavSusc} and \eqref{eq_MecSusc}, respectively.

We find that the annihilation operators of the output light field in Fourier space is
\begin{align} \label{eq_outputLadder}
    a_\rmmath{out}(\omega) &= \frac{x_\rmmath{c}^\rmmath{out}(\omega) + \i p_\rmmath{c}^\rmmath{out}(\omega)}{\sqrt{2}} \nonumber\\
    &= \mathcal{S}(\omega)a_\rmmath{in}(\omega) -\frac{\i\mathcal{T}(\omega)}{\sqrt{n_\mathrm{th} +1/2}} \xi(\omega) \nonumber\\
    &\quad + \i \mathcal{B}(\omega) \left( a_\rmmath{in}(\omega)+     a_\rmmath{in}^\dagger(-\omega) \right)
\end{align}
and $a_\rmmath{out}^\dagger(\omega) \coloneqq \left( a_\rmmath{out}(\omega) \right)^\dagger$. We used the relations between quadratures and ladder operators, \eref{eq_quadToLadder}, in Fourier space.

\subsection{Exact formula for the EPR-variance}\label{sec_exactEPRformula}
The mode operators, defined in \eref{eq_modeOp_time}, become a convolution in frequency domain
\begin{align}\label{eq_modeOp_freq}
    r_{i} &= \int_{-\infty}^\infty \rmmath{d}\omega\, f_{i}(-\omega) a_\rmmath{out}(\omega)
\end{align}
as a consequence of Plancherel's theorem.

Four different correlators appear in the expression of the EPR-variance in terms of ladder operators, \eref{eq_EPRvar_ladderOp}: $\langle r_j^\dagger\, r_j\rangle$, $ \langle r_j\, r_j^\dagger \rangle$, $ \langle r_j^\dagger\, r_k^\dagger  \rangle$ and $ \langle r_j\, r_k \rangle$; with $j,k=\rmmath{E,L}$ and $k\neq j$. We write them as frequency integrals using \eref{eq_modeOp_freq}
\begin{align}\label{eq_EPRcorrelators}
    \left\langle r_j^{(\dagger)} r_k^{(\dagger)} \right\rangle &= \iint^\infty_{-\infty} \rmmath{d}\omega\, \rmmath{d}\omega'\, f_j^{(*)}(-\omega) f_k^{(*)}(-\omega') \nonumber\\
    &\qquad \times \left\langle a_\rmmath{out}^{(\dagger)}(\omega) a_\rmmath{out}^{(\dagger)}(\omega') \right\rangle
\end{align}
where the superscripts $(\dagger)$ and $(*)$ indicate presence or absence of Hermitian and complex conjugation, respectively. One keeps track of the factor $e^{\i\phi}$ in \eref{eq_EPRvar_ladderOp} with the following replacement convention $f_\rmmath{L}(\omega) \rightarrow f_\rmmath{L}(\omega)e^{\i\phi}$.

We find the expressions of the four correlators above from \eref{eq_EPRcorrelators} using \eref{eq_outputLadder}, the Markovian bath's correlators \esref{eq_mechBath_correl} and \eqref{eq_lightBath_correl} in frequency space, and the shorthand notations and symmetries of \esref{eq_noiseFactors}. The thermal noise integrals need to be combined appropriately to take the from of \esref{eq_mechBath_correl} (in frequency space). Each correlator at this point is a single frequency integral and the EPR-variance, \eref{eq_EPRvar_ladderOp}, can be written as the following explicitly real expression
\begin{widetext}
\begin{equation}\label{eq_exactEPR}
\begin{split}
    \EPR &= \sum_{j=\rmmath{E,L}} \left[ \int_{-\infty}^\infty \rmmath{d}\omega\,f_j(\omega)f_j^*(\omega) \left( \frac{1}{2}\mathcal{S}(\omega) \mathcal{S}(-\omega) + \mathcal{B}(\omega)\mathcal{B}(-\omega) + \mathcal{T}(\omega)\mathcal{T}(-\omega) \right) + \rmmath{c.c.} \right]
    \\
    & - \sum_{\substack{j,k=\rmmath{E,L}\\k\neq j}} \left[ \int_{-\infty}^\infty \rmmath{d}\omega\,f_j(-\omega)f_k(\omega)\, \Big( \mathcal{B}(\omega)\mathcal{B}(-\omega) + \mathcal{T}(\omega)\mathcal{T}(-\omega) \Big) + \rmmath{c.c.} \right]
    \\
    & + \sum_{j=\rmmath{E,L}} \left[ \int_{-\infty}^\infty \rmmath{d}\omega\,f_j(\omega)f_j^*(\omega) \i \mathcal{S}(-\omega)\mathcal{B}(\omega) + \rmmath{c.c.} \right]
    \\
    & + \sum_{\substack{j,k=\rmmath{E,L}\\k\neq j}} \left[ \int_{-\infty}^\infty \rmmath{d}\omega\,f_j(-\omega)f_k(\omega)\, \i\mathcal{S}(-\omega) \mathcal{B}(\omega) + \rmmath{c.c.} \right].
\end{split}
\end{equation}
\end{widetext}
This relation is exact if the drive is not detuned with respect to the cavity. It also reveals how the driving noises couple to each other and to the mode functions: $\mathcal{S}(\omega) \mathcal{S}(-\omega)$, $\mathcal{B}(\omega)\mathcal{B}(-\omega)$ and $\mathcal{T}(\omega)\mathcal{T}(-\omega)$ come from the auto-correlations of the noises, while $\mathcal{S}(-\omega) \mathcal{B}(\omega)$ comes from the correlation between shot noise and back-action noise. The correlators are overlapped either with $f_jf_j^*$ -- we call them \emph{intra-mode} integrals -- or with $f_jf_k$ -- which we call \emph{inter-mode} integrals.

So far, we did not use that $f_\rmmath{E}$ and $f_\rmmath{L}$ are time-ordered and non-overlapping, therefore \eref{eq_exactEPR} is valid for arbitrary mode functions satisfying the orthonormality constraint, \eref{eq_orthoNormalisation}. Moreover, \eref{eq_exactEPR} takes a compact matrix form. Indeed, notice that $\mathcal{S}(\omega) \mathcal{S}(-\omega) = 1$, and call
\begin{subequations}
\begin{align}
D(\omega) &\coloneqq  \mathcal{B}(\omega)\mathcal{B}(-\omega) + \mathcal{T}(\omega)\mathcal{T}(-\omega) \nonumber\\
&\ = 4 g^2 |\chi_\rmmath{opt}(\omega)|^2  |\chi_\rmmath{m}(\omega)|^2 \big[g^2 |\chi_\rmmath{opt}(\omega)|^2\nonumber\\
&\quad\ + \gamma_\rmmath{m} (n_\mathrm{th} + 1/2) \big], \\
P(\omega) &\coloneqq  \mathcal{S}(-\omega)\mathcal{B}(\omega) \nonumber\\
&\ = 2g^2 \mathcal{S}(-\omega) \chi_\rmmath{opt}^2(\omega) \chi_\rmmath{m}(\omega)
\end{align}
\end{subequations}
with $D^*(\omega)=D(-\omega) = D(\omega)$ (i.e., real) and $P^*(\omega)=P(-\omega)$. Upon expanding the sums in \eref{eq_exactEPR}, making sure the argument of $f_\rmmath{L}$ is $-\omega$, one obtains the compact matrix expression
\begin{equation}\label{eq_EPRmatForm}
    \EPR = 2+ \int_{-\infty}^\infty \rmmath{d}\omega\, v^\dagger(\omega) M(\omega) v(\omega)
\end{equation}
where $v(\omega) \coloneqq \big(f_\mathrm{E}(\omega), f_\mathrm{L}^*(-\omega), f_\mathrm{E}^*(\omega), f_\mathrm{L}(-\omega) \big)^T$. The matrix $M$ is Hermitian as
\begin{equation}\label{eq_matM}
    M \coloneqq
    \begin{tiny}
    \begin{pmatrix}
        D-P_I & -D-\i P_R & 0 & 0\\
        -D+\i P_R & D+P_I & 0 & 0\\
         0 & 0 & D-P_I & -D+\i P_R\\
         0 & 0 & -D-\i P_R & D+ P_I
    \end{pmatrix}
   \end{tiny}
\end{equation}
with $P_R \coloneqq \rmmath{Re}[P]$ and $P_I \coloneqq \rmmath{Im}[P]$ being the real and imaginary parts of $P$, respectively.

It is interesting to look at the problem of minimising \eref{eq_EPRmatForm} under the constraints that the mode functions are time-ordered -- necessary to infer optomechanical entanglement based on light modes entanglement -- and orthonormal, \eref{eq_orthoNormalisation}. Titchmarsh's theorem \cite{Titchmarsh, Nussenzveig_72} states that the causality constraint is equivalent to mode functions determined by their real parts only. Therefore, the constraints can be formulated in a quadratic form such that the quadratic minimisation problem can be solved in polynomial times with linear programming methods, according to \cite{qcqp}. We were not successful with a simple and naive approach to perform the minimisation this way. We attempted to solve it as an eigenvalue problem but were not successful either: the main issues were the wide parameter spread (ratio between largest and smallest scale is $10^8$) and the form of the normalisation constraint. Moreover, optimising the mode functions to minimize $\EPR$ is sub-optimal but we expect that repeating our analysis for Duan's criterion \cite{Duan_2000} in its actual form or even a general witness that is linear in covariance matrices \cite{Hyllus_2006} is a simple generalisation.

\subsection{Approximate formula for the EPR-variance}\label{sec_DerivEPRformula}
Under the assumption that $\kappa \gg \omega_\rmmath{m} \gg \gamma_\rmmath{m}$ motivated in \sref{sec_tempModes}, we made the approximations of \esref{eq_approximatedFunctions}. We update the noise pre-factors of \esref{eq_noiseFactors} accordingly
\begin{subequations}\label{eq_ApproxSnBaTh}
\begin{align}
    \mathcal{S}(\omega) &\approx 1,\\
    \mathcal{B}(\omega) &\approx 2 \GammaRO \chi_\rmmath{m}(\omega),\\
    \mathcal{T}(\omega) &\approx 2 \sqrt{\GammaRO \Gamma_\rmmath{th}}\chi_\rmmath{m}(\omega)
\end{align}
\end{subequations}
such that $\mathcal{S}^*= \mathcal{S}$, $\mathcal{B}^*(\omega)= \mathcal{B}(-\omega)$ and $\mathcal{T}^*(\omega) = \mathcal{T}(-\omega)$. We defined $\GammaRO$ and $\Gamma_\rmmath{th}$ in \esref{eq_GammaRO} and \eqref{eq_Gammath}, respectively. These approximations simplify the terms in the curly brackets in \eref{eq_exactEPR}: the remaining $\omega$ dependence is $|\chi_\text{m}|^2$ or $\chi_\text{m}$.

The mode functions of \esref{eq_modeFunct_time} are Lorentzians in frequency space
\begin{equation}\label{eq_modeFunct_freq}
    f_\rmmath{E/L}(\omega) = N_\rmmath{E/L} \frac{e^{\mp \i \omega T_\rmmath{sep}/2}}{\omega - \omega_\rmmath{E/L}}
\end{equation}
with $\omega_\rmmath{E/L} \coloneqq \mp (\omega_\rmmath{m} + \i \Gamma)$ (minus signs associated to index E) and
\begin{equation}
N_\rmmath{E/L} = \mp \i \sqrt{\frac{\Gamma}{\pi}} e^{\i \omega_\rmmath{m} T_\rmmath{sep}/2}
\end{equation}
is the normalization factor.

The integrands in \eref{eq_exactEPR} are thus products of the mode functions, \eref{eq_modeFunct_freq}, with the mechanical susceptibility, \eref{eq_MecSuscApprox}, or its modulus square, \eref{eq_MecSuscApprox_modSquare}. Terms like $(\omega - \omega_\rmmath{E})^{-1} (\omega - \omega_-)^{-1}$ and $(\omega - \omega_\rmmath{L})^{-1} (\omega - \omega_+)^{-1}$ are resonant and the next largest terms scale relatively like $O\left(\Gamma \sqrt{n_\rmmath{th}(\Cq +1)}/\omega_\rmmath{m}\right)$. As a further (purely technical) simplification we keep only the resonant terms and require that $\Gamma \ll \omega_\rmmath{m}/\sqrt{n_\rmmath{th}(C+1)}$; this is the origin of \eref{eq_approxGamma}. The thus approximated four integrals of \eref{eq_exactEPR} can be evaluated with the residue theorem, we find [keeping the line structure of \eref{eq_exactEPR}]
\begin{equation}\label{eq_EPRformula_unitEff}
\begin{split}
    \EPR &=2+ 8(\GammaRO + \Gamma_\text{th}) \frac{1}{\gamma_\rmmath{m}(\Gamma + \frac{\gamma_\rmmath{m}}{2})}\\
    &\quad -8(\GammaRO + \Gamma_\text{th})\frac{\Gamma e^{-\gamma_\rmmath{m} T_\rmmath{sep}/2}}{\gamma_\rmmath{m}(\Gamma + \frac{\gamma_\rmmath{m}}{2})^2} \,\cos \phi\\
    &\quad + 0\\
    &\quad -4\GammaRO \frac{\Gamma e^{-\gamma_\rmmath{m} T_\rmmath{sep}/2}}{(\Gamma + \frac{\gamma_\rmmath{m}}{2})^2} \,\cos \phi.
\end{split}
\end{equation}
This concludes the derivation of the approximate formula of the EPR-variance, for ideal detection ($\eta=1$), in the regime $\kappa \gg \omega_\rmmath{m} \gg \gamma_\rmmath{m}$, $\Gamma \ll \omega_\rmmath{m}/\sqrt{n_\rmmath{th}(C+1)}$ and $\delta =0$. It is \eref{eq_EPRformula} evaluated at $\eta=1$ (called $\left.\EPR\right|_{\eta=1}$ below).

Finally, we model inefficient detection (passive losses) by a beam-splitter of transitivity $\eta$ right before an ideal (efficiency 1) detector. In the detected channel the field operator is
\begin{equation}
    a_\rmmath{out}^\rmmath{meas} = \sqrt{\eta} a_\rmmath{out} + \sqrt{1-\eta} a_\rmmath{shot}
\end{equation}
in which $a_\rmmath{out}$ is the cavity output operator as in \eref{eq_outputLadder}, and $a_\rmmath{shot}$ is the operator of shot noise that entered the free port of the beam-splitter. The correlators appearing in $\EPR$, \eref{eq_EPRcorrelators}, factorize because $a_\rmmath{out}$ and $a_\rmmath{shot}$ are uncorrelated. Using that the intra-pulse auto-correlators of shot noise equals 2, we find that the EPR-variance accounting for detection inefficiency, \eref{eq_EPRformula}, is given by
\begin{equation}
     \EPR = \left.\eta \EPR\right|_{\eta=1} + (1-\eta)2.
\end{equation}
We have denoted here with $\left.\EPR\right|_{\eta=1}$ the expression for the EPR-variance for unit detection efficiency derived in this appendix, \eref{eq_EPRformula_unitEff}.

\bibliography{biblio_arxiv}

\begin{thebibliography}{63}%
\makeatletter
\providecommand \@ifxundefined [1]{%
 \@ifx{#1\undefined}
}%
\providecommand \@ifnum [1]{%
 \ifnum #1\expandafter \@firstoftwo
 \else \expandafter \@secondoftwo
 \fi
}%
\providecommand \@ifx [1]{%
 \ifx #1\expandafter \@firstoftwo
 \else \expandafter \@secondoftwo
 \fi
}%
\providecommand \natexlab [1]{#1}%
\providecommand \enquote  [1]{``#1''}%
\providecommand \bibnamefont  [1]{#1}%
\providecommand \bibfnamefont [1]{#1}%
\providecommand \citenamefont [1]{#1}%
\providecommand \href@noop [0]{\@secondoftwo}%
\providecommand \href [0]{\begingroup \@sanitize@url \@href}%
\providecommand \@href[1]{\@@startlink{#1}\@@href}%
\providecommand \@@href[1]{\endgroup#1\@@endlink}%
\providecommand \@sanitize@url [0]{\catcode `\\12\catcode `\$12\catcode
  `\&12\catcode `\#12\catcode `\^12\catcode `\_12\catcode `\%12\relax}%
\providecommand \@@startlink[1]{}%
\providecommand \@@endlink[0]{}%
\providecommand \url  [0]{\begingroup\@sanitize@url \@url }%
\providecommand \@url [1]{\endgroup\@href {#1}{\urlprefix }}%
\providecommand \urlprefix  [0]{URL }%
\providecommand \Eprint [0]{\href }%
\providecommand \doibase [0]{http://dx.doi.org/}%
\providecommand \selectlanguage [0]{\@gobble}%
\providecommand \bibinfo  [0]{\@secondoftwo}%
\providecommand \bibfield  [0]{\@secondoftwo}%
\providecommand \translation [1]{[#1]}%
\providecommand \BibitemOpen [0]{}%
\providecommand \bibitemStop [0]{}%
\providecommand \bibitemNoStop [0]{.\EOS\space}%
\providecommand \EOS [0]{\spacefactor3000\relax}%
\providecommand \BibitemShut  [1]{\csname bibitem#1\endcsname}%
\let\auto@bib@innerbib\@empty
\bibitem [{\citenamefont {Teufel}\ \emph {et~al.}(2011)\citenamefont {Teufel},
  \citenamefont {Donner}, \citenamefont {Li}, \citenamefont {Harlow},
  \citenamefont {Allman}, \citenamefont {Cicak}, \citenamefont {Sirois},
  \citenamefont {Whittaker}, \citenamefont {Lehnert},\ and\ \citenamefont
  {Simmonds}}]{teufel_sideband_2011}%
  \BibitemOpen
  \bibfield  {author} {\bibinfo {author} {\bibfnamefont {J.~D.}\ \bibnamefont
  {Teufel}}, \bibinfo {author} {\bibfnamefont {T.}~\bibnamefont {Donner}},
  \bibinfo {author} {\bibfnamefont {D.}~\bibnamefont {Li}}, \bibinfo {author}
  {\bibfnamefont {J.~W.}\ \bibnamefont {Harlow}}, \bibinfo {author}
  {\bibfnamefont {M.~S.}\ \bibnamefont {Allman}}, \bibinfo {author}
  {\bibfnamefont {K.}~\bibnamefont {Cicak}}, \bibinfo {author} {\bibfnamefont
  {A.~J.}\ \bibnamefont {Sirois}}, \bibinfo {author} {\bibfnamefont {J.~D.}\
  \bibnamefont {Whittaker}}, \bibinfo {author} {\bibfnamefont {K.~W.}\
  \bibnamefont {Lehnert}}, \ and\ \bibinfo {author} {\bibfnamefont {R.~W.}\
  \bibnamefont {Simmonds}},\ }\href {\doibase 10.1038/nature10261} {\bibfield
  {journal} {\bibinfo  {journal} {Nature}\ }\textbf {\bibinfo {volume} {475}},\
  \bibinfo {pages} {359} (\bibinfo {year} {2011})}\BibitemShut {NoStop}%
\bibitem [{\citenamefont {Chan}\ \emph {et~al.}(2011)\citenamefont {Chan},
  \citenamefont {Alegre}, \citenamefont {Safavi-Naeini}, \citenamefont {Hill},
  \citenamefont {Krause}, \citenamefont {Groeblacher}, \citenamefont
  {Aspelmeyer},\ and\ \citenamefont {Painter}}]{Painter_2011}%
  \BibitemOpen
  \bibfield  {author} {\bibinfo {author} {\bibfnamefont {J.}~\bibnamefont
  {Chan}}, \bibinfo {author} {\bibfnamefont {T.}~\bibnamefont {Alegre}},
  \bibinfo {author} {\bibfnamefont {A.}~\bibnamefont {Safavi-Naeini}}, \bibinfo
  {author} {\bibfnamefont {J.}~\bibnamefont {Hill}}, \bibinfo {author}
  {\bibfnamefont {A.}~\bibnamefont {Krause}}, \bibinfo {author} {\bibfnamefont
  {S.}~\bibnamefont {Groeblacher}}, \bibinfo {author} {\bibfnamefont
  {M.}~\bibnamefont {Aspelmeyer}}, \ and\ \bibinfo {author} {\bibfnamefont
  {O.}~\bibnamefont {Painter}},\ }\href {\doibase 10.1038/nature10461}
  {\bibfield  {journal} {\bibinfo  {journal} {Nature}\ }\textbf {\bibinfo
  {volume} {478}},\ \bibinfo {pages} {89} (\bibinfo {year} {2011})}\BibitemShut
  {NoStop}%
\bibitem [{\citenamefont {Brooks}\ \emph {et~al.}(2012)\citenamefont {Brooks},
  \citenamefont {Botter}, \citenamefont {Schreppler}, \citenamefont {Purdy},
  \citenamefont {Brahms},\ and\ \citenamefont
  {Stamper-Kurn}}]{Stamper-Kurn_2012}%
  \BibitemOpen
  \bibfield  {author} {\bibinfo {author} {\bibfnamefont {D.~W.~C.}\
  \bibnamefont {Brooks}}, \bibinfo {author} {\bibfnamefont {T.}~\bibnamefont
  {Botter}}, \bibinfo {author} {\bibfnamefont {S.}~\bibnamefont {Schreppler}},
  \bibinfo {author} {\bibfnamefont {T.~P.}\ \bibnamefont {Purdy}}, \bibinfo
  {author} {\bibfnamefont {N.}~\bibnamefont {Brahms}}, \ and\ \bibinfo {author}
  {\bibfnamefont {D.~M.}\ \bibnamefont {Stamper-Kurn}},\ }\href {\doibase
  10.1038/nature11325} {\bibfield  {journal} {\bibinfo  {journal} {Nature}\
  }\textbf {\bibinfo {volume} {488}},\ \bibinfo {pages} {476} (\bibinfo {year}
  {2012})}\BibitemShut {NoStop}%
\bibitem [{\citenamefont {Safavi-Naeini}\ \emph {et~al.}(2013)\citenamefont
  {Safavi-Naeini}, \citenamefont {Gröblacher}, \citenamefont {Hill},
  \citenamefont {Chan}, \citenamefont {Aspelmeyer},\ and\ \citenamefont
  {Painter}}]{Safavi-Naeini_2013}%
  \BibitemOpen
  \bibfield  {author} {\bibinfo {author} {\bibfnamefont {A.~H.}\ \bibnamefont
  {Safavi-Naeini}}, \bibinfo {author} {\bibfnamefont {S.}~\bibnamefont
  {Gröblacher}}, \bibinfo {author} {\bibfnamefont {J.~T.}\ \bibnamefont
  {Hill}}, \bibinfo {author} {\bibfnamefont {J.}~\bibnamefont {Chan}}, \bibinfo
  {author} {\bibfnamefont {M.}~\bibnamefont {Aspelmeyer}}, \ and\ \bibinfo
  {author} {\bibfnamefont {O.}~\bibnamefont {Painter}},\ }\href {\doibase
  10.1038/nature12307} {\bibfield  {journal} {\bibinfo  {journal} {Nature}\
  }\textbf {\bibinfo {volume} {500}},\ \bibinfo {pages} {185} (\bibinfo {year}
  {2013})}\BibitemShut {NoStop}%
\bibitem [{\citenamefont {Purdy}\ \emph
  {et~al.}(2013{\natexlab{a}})\citenamefont {Purdy}, \citenamefont {Yu},
  \citenamefont {Peterson}, \citenamefont {Kampel},\ and\ \citenamefont
  {Regal}}]{Purdy2013a}%
  \BibitemOpen
  \bibfield  {author} {\bibinfo {author} {\bibfnamefont {T.~P.}\ \bibnamefont
  {Purdy}}, \bibinfo {author} {\bibfnamefont {P.-L.}\ \bibnamefont {Yu}},
  \bibinfo {author} {\bibfnamefont {R.~W.}\ \bibnamefont {Peterson}}, \bibinfo
  {author} {\bibfnamefont {N.~S.}\ \bibnamefont {Kampel}}, \ and\ \bibinfo
  {author} {\bibfnamefont {C.~A.}\ \bibnamefont {Regal}},\ }\href {\doibase
  10.1103/PhysRevX.3.031012} {\bibfield  {journal} {\bibinfo  {journal} {Phys.
  Rev. X}\ }\textbf {\bibinfo {volume} {3}},\ \bibinfo {pages} {031012}
  (\bibinfo {year} {2013}{\natexlab{a}})}\BibitemShut {NoStop}%
\bibitem [{\citenamefont {Barzanjeh}\ \emph {et~al.}(2019)\citenamefont
  {Barzanjeh}, \citenamefont {Redchenko}, \citenamefont {Redchenko},
  \citenamefont {Peruzzo}, \citenamefont {Wulf}, \citenamefont {Lewis},
  \citenamefont {Arnold},\ and\ \citenamefont {Fink}}]{Barzanjeh_2019}%
  \BibitemOpen
  \bibfield  {author} {\bibinfo {author} {\bibfnamefont {S.}~\bibnamefont
  {Barzanjeh}}, \bibinfo {author} {\bibfnamefont {E.~S.}\ \bibnamefont
  {Redchenko}}, \bibinfo {author} {\bibfnamefont {E.~S.}\ \bibnamefont
  {Redchenko}}, \bibinfo {author} {\bibfnamefont {M.}~\bibnamefont {Peruzzo}},
  \bibinfo {author} {\bibfnamefont {M.}~\bibnamefont {Wulf}}, \bibinfo {author}
  {\bibfnamefont {D.~P.}\ \bibnamefont {Lewis}}, \bibinfo {author}
  {\bibfnamefont {G.}~\bibnamefont {Arnold}}, \ and\ \bibinfo {author}
  {\bibfnamefont {J.~M.}\ \bibnamefont {Fink}},\ }\href {\doibase
  10.1038/s41586-019-1320-2} {\bibfield  {journal} {\bibinfo  {journal}
  {Nature}\ }\textbf {\bibinfo {volume} {570}},\ \bibinfo {pages} {480}
  (\bibinfo {year} {2019})}\BibitemShut {NoStop}%
\bibitem [{\citenamefont {Chen}\ \emph {et~al.}(2020)\citenamefont {Chen},
  \citenamefont {Rossi}, \citenamefont {Mason},\ and\ \citenamefont
  {Schliesser}}]{Schliesser_2020}%
  \BibitemOpen
  \bibfield  {author} {\bibinfo {author} {\bibfnamefont {J.}~\bibnamefont
  {Chen}}, \bibinfo {author} {\bibfnamefont {M.}~\bibnamefont {Rossi}},
  \bibinfo {author} {\bibfnamefont {D.}~\bibnamefont {Mason}}, \ and\ \bibinfo
  {author} {\bibfnamefont {A.}~\bibnamefont {Schliesser}},\ }\href
  {https://doi.org/10.1038/s41467-020-14768-1} {\bibfield  {journal} {\bibinfo
  {journal} {Nat Commun}\ }\textbf {\bibinfo {volume} {11}} (\bibinfo {year}
  {2020})}\BibitemShut {NoStop}%
\bibitem [{\citenamefont {Ockeloen-Korppi}\ \emph {et~al.}(2018)\citenamefont
  {Ockeloen-Korppi}, \citenamefont {Damskägg}, \citenamefont {Pirkkalainen},
  \citenamefont {Asjad}, \citenamefont {Clerk}, \citenamefont {Massel},
  \citenamefont {Woolley},\ and\ \citenamefont
  {Sillanpää}}]{Ockeloen-Korppi2018}%
  \BibitemOpen
  \bibfield  {author} {\bibinfo {author} {\bibfnamefont {C.~F.}\ \bibnamefont
  {Ockeloen-Korppi}}, \bibinfo {author} {\bibfnamefont {E.}~\bibnamefont
  {Damskägg}}, \bibinfo {author} {\bibfnamefont {J.-M.}\ \bibnamefont
  {Pirkkalainen}}, \bibinfo {author} {\bibfnamefont {M.}~\bibnamefont {Asjad}},
  \bibinfo {author} {\bibfnamefont {A.~A.}\ \bibnamefont {Clerk}}, \bibinfo
  {author} {\bibfnamefont {F.}~\bibnamefont {Massel}}, \bibinfo {author}
  {\bibfnamefont {M.~J.}\ \bibnamefont {Woolley}}, \ and\ \bibinfo {author}
  {\bibfnamefont {M.~A.}\ \bibnamefont {Sillanpää}},\ }\href {\doibase
  10.1038/s41586-018-0038-x} {\bibfield  {journal} {\bibinfo  {journal}
  {Nature}\ }\textbf {\bibinfo {volume} {556}},\ \bibinfo {pages} {478}
  (\bibinfo {year} {2018})}\BibitemShut {NoStop}%
\bibitem [{\citenamefont {Riedinger}\ \emph {et~al.}(2018)\citenamefont
  {Riedinger}, \citenamefont {Marinković}, \citenamefont {Löschnauer},
  \citenamefont {Aspelmeyer}, \citenamefont {Hong},\ and\ \citenamefont
  {Gröblacher}}]{Riedinger2018}%
  \BibitemOpen
  \bibfield  {author} {\bibinfo {author} {\bibfnamefont {A.}~\bibnamefont
  {Riedinger}, \bibfnamefont {Ralfand~Wallucks}}, \bibinfo {author}
  {\bibfnamefont {I.}~\bibnamefont {Marinković}}, \bibinfo {author}
  {\bibfnamefont {C.}~\bibnamefont {Löschnauer}}, \bibinfo {author}
  {\bibfnamefont {M.}~\bibnamefont {Aspelmeyer}}, \bibinfo {author}
  {\bibfnamefont {S.}~\bibnamefont {Hong}}, \ and\ \bibinfo {author}
  {\bibfnamefont {S.}~\bibnamefont {Gröblacher}},\ }\href {\doibase
  10.1038/s41586-018-0036-z} {\bibfield  {journal} {\bibinfo  {journal}
  {Nature}\ }\textbf {\bibinfo {volume} {556}},\ \bibinfo {pages} {473}
  (\bibinfo {year} {2018})}\BibitemShut {NoStop}%
\bibitem [{\citenamefont {Purdy}\ \emph
  {et~al.}(2013{\natexlab{b}})\citenamefont {Purdy}, \citenamefont {Peterson},\
  and\ \citenamefont {Regal}}]{Purdy2013}%
  \BibitemOpen
  \bibfield  {author} {\bibinfo {author} {\bibfnamefont {T.~P.}\ \bibnamefont
  {Purdy}}, \bibinfo {author} {\bibfnamefont {R.~W.}\ \bibnamefont {Peterson}},
  \ and\ \bibinfo {author} {\bibfnamefont {C.~A.}\ \bibnamefont {Regal}},\
  }\href {\doibase 10.1126/science.1231282} {\bibfield  {journal} {\bibinfo
  {journal} {Science}\ }\textbf {\bibinfo {volume} {339}},\ \bibinfo {pages}
  {801} (\bibinfo {year} {2013}{\natexlab{b}})}\BibitemShut {NoStop}%
\bibitem [{\citenamefont {Lei}\ \emph {et~al.}(2016)\citenamefont {Lei},
  \citenamefont {Weinstein}, \citenamefont {Suh}, \citenamefont {Wollman},
  \citenamefont {Kronwald}, \citenamefont {Marquardt}, \citenamefont {Clerk},\
  and\ \citenamefont {Schwab}}]{Schwab_2016}%
  \BibitemOpen
  \bibfield  {author} {\bibinfo {author} {\bibfnamefont {C.~U.}\ \bibnamefont
  {Lei}}, \bibinfo {author} {\bibfnamefont {A.~J.}\ \bibnamefont {Weinstein}},
  \bibinfo {author} {\bibfnamefont {J.}~\bibnamefont {Suh}}, \bibinfo {author}
  {\bibfnamefont {E.~E.}\ \bibnamefont {Wollman}}, \bibinfo {author}
  {\bibfnamefont {A.}~\bibnamefont {Kronwald}}, \bibinfo {author}
  {\bibfnamefont {F.}~\bibnamefont {Marquardt}}, \bibinfo {author}
  {\bibfnamefont {A.~A.}\ \bibnamefont {Clerk}}, \ and\ \bibinfo {author}
  {\bibfnamefont {K.~C.}\ \bibnamefont {Schwab}},\ }\href {\doibase
  10.1103/PhysRevLett.117.100801} {\bibfield  {journal} {\bibinfo  {journal}
  {Phys. Rev. Lett.}\ }\textbf {\bibinfo {volume} {117}},\ \bibinfo {pages}
  {100801} (\bibinfo {year} {2016})}\BibitemShut {NoStop}%
\bibitem [{\citenamefont {Clark}\ \emph {et~al.}(2017)\citenamefont {Clark},
  \citenamefont {Lecocq}, \citenamefont {Simmonds}, \citenamefont {Aumentado},\
  and\ \citenamefont {Teufel}}]{Teufel_2017}%
  \BibitemOpen
  \bibfield  {author} {\bibinfo {author} {\bibfnamefont {J.~B.}\ \bibnamefont
  {Clark}}, \bibinfo {author} {\bibfnamefont {F.}~\bibnamefont {Lecocq}},
  \bibinfo {author} {\bibfnamefont {R.~W.}\ \bibnamefont {Simmonds}}, \bibinfo
  {author} {\bibfnamefont {J.}~\bibnamefont {Aumentado}}, \ and\ \bibinfo
  {author} {\bibfnamefont {J.~D.}\ \bibnamefont {Teufel}},\ }\href {\doibase
  10.1038/nature20604} {\bibfield  {journal} {\bibinfo  {journal} {Nature}\
  }\textbf {\bibinfo {volume} {541}},\ \bibinfo {pages} {191} (\bibinfo {year}
  {2017})}\BibitemShut {NoStop}%
\bibitem [{\citenamefont {Shomroni}\ \emph {et~al.}(2019)\citenamefont
  {Shomroni}, \citenamefont {Malz}, \citenamefont {Nunnenkamp},\ and\
  \citenamefont {Kippenberg}}]{Shomroni2019}%
  \BibitemOpen
  \bibfield  {author} {\bibinfo {author} {\bibfnamefont {L.}~\bibnamefont
  {Shomroni}, \bibfnamefont {ItayQiu}}, \bibinfo {author} {\bibfnamefont
  {D.}~\bibnamefont {Malz}}, \bibinfo {author} {\bibfnamefont {A.}~\bibnamefont
  {Nunnenkamp}}, \ and\ \bibinfo {author} {\bibfnamefont {T.~J.}\ \bibnamefont
  {Kippenberg}},\ }\href {\doibase 10.1038/s41467-019-10024-3} {\bibfield
  {journal} {\bibinfo  {journal} {Nature Comm.}\ }\textbf {\bibinfo {volume}
  {10}},\ \bibinfo {pages} {2086} (\bibinfo {year} {2019})}\BibitemShut
  {NoStop}%
\bibitem [{\citenamefont {Palomaki}\ \emph {et~al.}(2013)\citenamefont
  {Palomaki}, \citenamefont {Teufel}, \citenamefont {Simmonds},\ and\
  \citenamefont {Lehnert}}]{Palomaki_2013}%
  \BibitemOpen
  \bibfield  {author} {\bibinfo {author} {\bibfnamefont {T.~A.}\ \bibnamefont
  {Palomaki}}, \bibinfo {author} {\bibfnamefont {J.~D.}\ \bibnamefont
  {Teufel}}, \bibinfo {author} {\bibfnamefont {R.~W.}\ \bibnamefont
  {Simmonds}}, \ and\ \bibinfo {author} {\bibfnamefont {K.~W.}\ \bibnamefont
  {Lehnert}},\ }\href {\doibase 10.1126/science.1244563} {\bibfield  {journal}
  {\bibinfo  {journal} {Science}\ }\textbf {\bibinfo {volume} {342}},\ \bibinfo
  {pages} {710} (\bibinfo {year} {2013})}\BibitemShut {NoStop}%
\bibitem [{\citenamefont {Marinkovi\ifmmode~\acute{c}\else \'{c}\fi{}}\ \emph
  {et~al.}(2018)\citenamefont {Marinkovi\ifmmode~\acute{c}\else \'{c}\fi{}},
  \citenamefont {Wallucks}, \citenamefont {Riedinger}, \citenamefont {Hong},
  \citenamefont {Aspelmeyer},\ and\ \citenamefont
  {Gr\"oblacher}}]{Groeblacher_2018}%
  \BibitemOpen
  \bibfield  {author} {\bibinfo {author} {\bibfnamefont {I.}~\bibnamefont
  {Marinkovi\ifmmode~\acute{c}\else \'{c}\fi{}}}, \bibinfo {author}
  {\bibfnamefont {A.}~\bibnamefont {Wallucks}}, \bibinfo {author}
  {\bibfnamefont {R.}~\bibnamefont {Riedinger}}, \bibinfo {author}
  {\bibfnamefont {S.}~\bibnamefont {Hong}}, \bibinfo {author} {\bibfnamefont
  {M.}~\bibnamefont {Aspelmeyer}}, \ and\ \bibinfo {author} {\bibfnamefont
  {S.}~\bibnamefont {Gr\"oblacher}},\ }\href {\doibase
  10.1103/PhysRevLett.121.220404} {\bibfield  {journal} {\bibinfo  {journal}
  {Phys. Rev. Lett.}\ }\textbf {\bibinfo {volume} {121}},\ \bibinfo {pages}
  {220404} (\bibinfo {year} {2018})}\BibitemShut {NoStop}%
\bibitem [{\citenamefont {Genes}\ \emph {et~al.}(2008)\citenamefont {Genes},
  \citenamefont {Mari}, \citenamefont {Tombesi},\ and\ \citenamefont
  {Vitali}}]{Genes_2008}%
  \BibitemOpen
  \bibfield  {author} {\bibinfo {author} {\bibfnamefont {C.}~\bibnamefont
  {Genes}}, \bibinfo {author} {\bibfnamefont {A.}~\bibnamefont {Mari}},
  \bibinfo {author} {\bibfnamefont {P.}~\bibnamefont {Tombesi}}, \ and\
  \bibinfo {author} {\bibfnamefont {D.}~\bibnamefont {Vitali}},\ }\href
  {\doibase 10.1103/physreva.78.032316} {\bibfield  {journal} {\bibinfo
  {journal} {Phys. Rev. A}\ }\textbf {\bibinfo {volume} {78}},\ \bibinfo
  {pages} {032316} (\bibinfo {year} {2008})}\BibitemShut {NoStop}%
\bibitem [{\citenamefont {Genes}\ \emph {et~al.}(2009)\citenamefont {Genes},
  \citenamefont {Mari}, \citenamefont {Vitali},\ and\ \citenamefont
  {Tombesi}}]{Genes_2009}%
  \BibitemOpen
  \bibfield  {author} {\bibinfo {author} {\bibfnamefont {C.}~\bibnamefont
  {Genes}}, \bibinfo {author} {\bibfnamefont {A.}~\bibnamefont {Mari}},
  \bibinfo {author} {\bibfnamefont {D.}~\bibnamefont {Vitali}}, \ and\ \bibinfo
  {author} {\bibfnamefont {P.}~\bibnamefont {Tombesi}},\ }\href {\doibase
  https://doi.org/10.1016/S1049-250X(09)57002-4} {\emph {\bibinfo {title}
  {Advances in Atomic Molecular and Optical Physics}}},\ \bibinfo {series}
  {Adv. At. Mol. Opt. Phys.}, Vol.~\bibinfo {volume} {57}\ (\bibinfo
  {publisher} {Academic Press},\ \bibinfo {year} {2009})\ pp.\ \bibinfo {pages}
  {33 -- 86}\BibitemShut {NoStop}%
\bibitem [{\citenamefont {Hofer}\ \emph {et~al.}(2011)\citenamefont {Hofer},
  \citenamefont {Wieczorek}, \citenamefont {Aspelmeyer},\ and\ \citenamefont
  {Hammerer}}]{Hofer_2011}%
  \BibitemOpen
  \bibfield  {author} {\bibinfo {author} {\bibfnamefont {S.~G.}\ \bibnamefont
  {Hofer}}, \bibinfo {author} {\bibfnamefont {W.}~\bibnamefont {Wieczorek}},
  \bibinfo {author} {\bibfnamefont {M.}~\bibnamefont {Aspelmeyer}}, \ and\
  \bibinfo {author} {\bibfnamefont {K.}~\bibnamefont {Hammerer}},\ }\href
  {\doibase 10.1103/physreva.84.052327} {\bibfield  {journal} {\bibinfo
  {journal} {Phys. Rev. A}\ }\textbf {\bibinfo {volume} {84}},\ \bibinfo
  {pages} {052327} (\bibinfo {year} {2011})}\BibitemShut {NoStop}%
\bibitem [{\citenamefont {Miao}\ \emph
  {et~al.}(2010{\natexlab{a}})\citenamefont {Miao}, \citenamefont {Danilishin},
  \citenamefont {Müller-Ebhardt},\ and\ \citenamefont {Chen}}]{Miao2}%
  \BibitemOpen
  \bibfield  {author} {\bibinfo {author} {\bibfnamefont {H.}~\bibnamefont
  {Miao}}, \bibinfo {author} {\bibfnamefont {S.}~\bibnamefont {Danilishin}},
  \bibinfo {author} {\bibfnamefont {H.}~\bibnamefont {Müller-Ebhardt}}, \ and\
  \bibinfo {author} {\bibfnamefont {Y.}~\bibnamefont {Chen}},\ }\href {\doibase
  10.1088/1367-2630/12/8/083032} {\bibfield  {journal} {\bibinfo  {journal}
  {New Journal of Physics}\ }\textbf {\bibinfo {volume} {12}},\ \bibinfo
  {pages} {083032} (\bibinfo {year} {2010}{\natexlab{a}})}\BibitemShut
  {NoStop}%
\bibitem [{\citenamefont {Miao}\ \emph
  {et~al.}(2010{\natexlab{b}})\citenamefont {Miao}, \citenamefont {Danilishin},
  \citenamefont {M\"uller-Ebhardt}, \citenamefont {Rehbein}, \citenamefont
  {Somiya},\ and\ \citenamefont {Chen}}]{Miao3}%
  \BibitemOpen
  \bibfield  {author} {\bibinfo {author} {\bibfnamefont {H.}~\bibnamefont
  {Miao}}, \bibinfo {author} {\bibfnamefont {S.}~\bibnamefont {Danilishin}},
  \bibinfo {author} {\bibfnamefont {H.}~\bibnamefont {M\"uller-Ebhardt}},
  \bibinfo {author} {\bibfnamefont {H.}~\bibnamefont {Rehbein}}, \bibinfo
  {author} {\bibfnamefont {K.}~\bibnamefont {Somiya}}, \ and\ \bibinfo {author}
  {\bibfnamefont {Y.}~\bibnamefont {Chen}},\ }\href {\doibase
  10.1103/PhysRevA.81.012114} {\bibfield  {journal} {\bibinfo  {journal} {Phys.
  Rev. A}\ }\textbf {\bibinfo {volume} {81}},\ \bibinfo {pages} {012114}
  (\bibinfo {year} {2010}{\natexlab{b}})}\BibitemShut {NoStop}%
\bibitem [{Mia(2012)}]{Miao4}%
  \BibitemOpen
  \href
  {https://www.tamagawa.jp/research/quantum/discourse/pdf/2013_ECSiaQIS_009_018_DMMC_Danilishin.pdf}
  {\emph {\bibinfo {title} {Optomechanical entanglement: How to prepare, verify
  and ”steer” a macroscopic mechanicalquantum state?}}}\ (\bibinfo
  {publisher} {Tamagawa University Quantum ICT Research Institute},\ \bibinfo
  {year} {2012})\BibitemShut {NoStop}%
\bibitem [{\citenamefont {Rossi}\ \emph {et~al.}(2019)\citenamefont {Rossi},
  \citenamefont {Mason}, \citenamefont {Chen},\ and\ \citenamefont
  {Schliesser}}]{rossi_observing_2019}%
  \BibitemOpen
  \bibfield  {author} {\bibinfo {author} {\bibfnamefont {M.}~\bibnamefont
  {Rossi}}, \bibinfo {author} {\bibfnamefont {D.}~\bibnamefont {Mason}},
  \bibinfo {author} {\bibfnamefont {J.}~\bibnamefont {Chen}}, \ and\ \bibinfo
  {author} {\bibfnamefont {A.}~\bibnamefont {Schliesser}},\ }\href {\doibase
  10.1103/PhysRevLett.123.163601} {\bibfield  {journal} {\bibinfo  {journal}
  {Physical Review Letters}\ }\textbf {\bibinfo {volume} {123}},\ \bibinfo
  {pages} {163601} (\bibinfo {year} {2019})}\BibitemShut {NoStop}%
\bibitem [{\citenamefont {Miao}\ \emph
  {et~al.}(2010{\natexlab{c}})\citenamefont {Miao}, \citenamefont
  {Danilishin},\ and\ \citenamefont {Chen}}]{Miao1}%
  \BibitemOpen
  \bibfield  {author} {\bibinfo {author} {\bibfnamefont {H.}~\bibnamefont
  {Miao}}, \bibinfo {author} {\bibfnamefont {S.}~\bibnamefont {Danilishin}}, \
  and\ \bibinfo {author} {\bibfnamefont {Y.}~\bibnamefont {Chen}},\ }\href
  {\doibase 10.1103/PhysRevA.81.052307} {\bibfield  {journal} {\bibinfo
  {journal} {Phys. Rev. A}\ }\textbf {\bibinfo {volume} {81}},\ \bibinfo
  {pages} {052307} (\bibinfo {year} {2010}{\natexlab{c}})}\BibitemShut
  {NoStop}%
\bibitem [{\citenamefont {von Neumann}(1955)}]{VonNeumann}%
  \BibitemOpen
  \bibfield  {author} {\bibinfo {author} {\bibfnamefont {J.}~\bibnamefont {von
  Neumann}},\ }\href@noop {} {\emph {\bibinfo {title} {Mathematical foundations
  of quantum mechanics}}}\ (\bibinfo  {publisher} {Princeton University
  Press},\ \bibinfo {year} {1955})\BibitemShut {NoStop}%
\bibitem [{\citenamefont {Zurek}(1981)}]{Zurek_81}%
  \BibitemOpen
  \bibfield  {author} {\bibinfo {author} {\bibfnamefont {W.}~\bibnamefont
  {Zurek}},\ }\href {\doibase 10.1103/PhysRevD.24.1516} {\bibfield  {journal}
  {\bibinfo  {journal} {Phys. Rev. D}\ }\textbf {\bibinfo {volume} {24}},\
  \bibinfo {pages} {1516} (\bibinfo {year} {1981})}\BibitemShut {NoStop}%
\bibitem [{\citenamefont {Zurek}(2003)}]{Zurek_03}%
  \BibitemOpen
  \bibfield  {author} {\bibinfo {author} {\bibfnamefont {W.}~\bibnamefont
  {Zurek}},\ }\href {\doibase 10.1103/RevModPhys.75.715} {\bibfield  {journal}
  {\bibinfo  {journal} {Rev. Mod. Phys.}\ }\textbf {\bibinfo {volume} {75}},\
  \bibinfo {pages} {715} (\bibinfo {year} {2003})}\BibitemShut {NoStop}%
\bibitem [{\citenamefont {Aspelmeyer}\ \emph {et~al.}(2014)\citenamefont
  {Aspelmeyer}, \citenamefont {Kippenberg},\ and\ \citenamefont
  {Marquardt}}]{OM_Review2014}%
  \BibitemOpen
  \bibfield  {author} {\bibinfo {author} {\bibfnamefont {M.}~\bibnamefont
  {Aspelmeyer}}, \bibinfo {author} {\bibfnamefont {T.~J.}\ \bibnamefont
  {Kippenberg}}, \ and\ \bibinfo {author} {\bibfnamefont {F.}~\bibnamefont
  {Marquardt}},\ }\href {\doibase 10.1103/RevModPhys.86.1391} {\bibfield
  {journal} {\bibinfo  {journal} {Rev. Mod. Phys.}\ }\textbf {\bibinfo {volume}
  {86}},\ \bibinfo {pages} {1391} (\bibinfo {year} {2014})}\BibitemShut
  {NoStop}%
\bibitem [{\citenamefont {Rabl}(2011)}]{Rabl2011}%
  \BibitemOpen
  \bibfield  {author} {\bibinfo {author} {\bibfnamefont {P.}~\bibnamefont
  {Rabl}},\ }\href {\doibase 10.1103/PhysRevLett.107.063601} {\bibfield
  {journal} {\bibinfo  {journal} {Phys. Rev. Lett.}\ }\textbf {\bibinfo
  {volume} {107}},\ \bibinfo {pages} {063601} (\bibinfo {year}
  {2011})}\BibitemShut {NoStop}%
\bibitem [{\citenamefont {Nunnenkamp}\ \emph {et~al.}(2011)\citenamefont
  {Nunnenkamp}, \citenamefont {B\o{}rkje},\ and\ \citenamefont
  {Girvin}}]{Nunnenkamp2011}%
  \BibitemOpen
  \bibfield  {author} {\bibinfo {author} {\bibfnamefont {A.}~\bibnamefont
  {Nunnenkamp}}, \bibinfo {author} {\bibfnamefont {K.}~\bibnamefont
  {B\o{}rkje}}, \ and\ \bibinfo {author} {\bibfnamefont {S.~M.}\ \bibnamefont
  {Girvin}},\ }\href {\doibase 10.1103/PhysRevLett.107.063602} {\bibfield
  {journal} {\bibinfo  {journal} {Phys. Rev. Lett.}\ }\textbf {\bibinfo
  {volume} {107}},\ \bibinfo {pages} {063602} (\bibinfo {year}
  {2011})}\BibitemShut {NoStop}%
\bibitem [{\citenamefont {Gardiner}\ and\ \citenamefont
  {Zoller}(2010)}]{Zoller_Gardiner2010}%
  \BibitemOpen
  \bibfield  {author} {\bibinfo {author} {\bibfnamefont {C.~W.}\ \bibnamefont
  {Gardiner}}\ and\ \bibinfo {author} {\bibfnamefont {P.}~\bibnamefont
  {Zoller}},\ }\href@noop {} {{\selectlanguage {english}\emph {\bibinfo {title}
  {Quantum noise}}}},\ \bibinfo {edition} {3rd}\ ed.\ (\bibinfo  {publisher}
  {Springer},\ \bibinfo {address} {Berlin},\ \bibinfo {year}
  {2010})\BibitemShut {NoStop}%
\bibitem [{\citenamefont {Caldeira}\ and\ \citenamefont
  {Leggett}(1983)}]{Caldeira_Leggett_83}%
  \BibitemOpen
  \bibfield  {author} {\bibinfo {author} {\bibfnamefont {A.}~\bibnamefont
  {Caldeira}}\ and\ \bibinfo {author} {\bibfnamefont {A.}~\bibnamefont
  {Leggett}},\ }\href {\doibase https://doi.org/10.1016/0378-4371(83)90013-4}
  {\bibfield  {journal} {\bibinfo  {journal} {Physica A}\ }\textbf {\bibinfo
  {volume} {121}},\ \bibinfo {pages} {587 } (\bibinfo {year}
  {1983})}\BibitemShut {NoStop}%
\bibitem [{Note1()}]{Note1}%
  \BibitemOpen
  \bibinfo {note} {The QLE can be either expressed with the Brownian noise
  affecting the mechanical momentum quadrature only, as in Eqs.~\protect
  \textup {\hbox {\mathsurround \z@ \protect \normalfont (\ignorespaces \ref
  {eq_QLEtime}\unskip \@@italiccorr )}}, or with the Brownian noise affecting
  both mechanical quadratures symmetrically. We performed the exact numerical
  computation based on the QLE with symmetric and asymmetric Brownian noise and
  we found that all the results presented in this work are the same in both
  cases, up to a few parts per million.}\BibitemShut {Stop}%
\bibitem [{Note2()}]{Note2}%
  \BibitemOpen
  \bibinfo {note} {The stability is most easily assessed by the Routh--Hurwitz
  criterion. Whenever the detuning is zero the dynamics is unconditionally
  stable \cite {Thesis_Hofer}.}\BibitemShut {Stop}%
\bibitem [{\citenamefont {Gardiner}\ and\ \citenamefont
  {Collett}(1985)}]{Gardiner_Collet_85}%
  \BibitemOpen
  \bibfield  {author} {\bibinfo {author} {\bibfnamefont {C.~W.}\ \bibnamefont
  {Gardiner}}\ and\ \bibinfo {author} {\bibfnamefont {M.~J.}\ \bibnamefont
  {Collett}},\ }\href {\doibase 10.1103/PhysRevA.31.3761} {\bibfield  {journal}
  {\bibinfo  {journal} {Phys. Rev. A}\ }\textbf {\bibinfo {volume} {31}},\
  \bibinfo {pages} {3761} (\bibinfo {year} {1985})}\BibitemShut {NoStop}%
\bibitem [{\citenamefont {Cubitt}\ \emph {et~al.}(2003)\citenamefont {Cubitt},
  \citenamefont {Verstraete}, \citenamefont {D{\"u}r},\ and\ \citenamefont
  {Cirac}}]{Cubitt_2003}%
  \BibitemOpen
  \bibfield  {author} {\bibinfo {author} {\bibfnamefont {T.~S.}\ \bibnamefont
  {Cubitt}}, \bibinfo {author} {\bibfnamefont {F.}~\bibnamefont {Verstraete}},
  \bibinfo {author} {\bibfnamefont {W.}~\bibnamefont {D{\"u}r}}, \ and\
  \bibinfo {author} {\bibfnamefont {J.~I.}\ \bibnamefont {Cirac}},\ }\href
  {\doibase 10.1103/physrevlett.91.037902} {\bibfield  {journal} {\bibinfo
  {journal} {Phys. Rev. Lett.}\ }\textbf {\bibinfo {volume} {91}},\ \bibinfo
  {pages} {037902} (\bibinfo {year} {2003})}\BibitemShut {NoStop}%
\bibitem [{\citenamefont {Mi\ifmmode~\check{s}\else \v{s}\fi{}ta}\ and\
  \citenamefont {Korolkova}(2008)}]{Korolkova_08}%
  \BibitemOpen
  \bibfield  {author} {\bibinfo {author} {\bibfnamefont {L.}~\bibnamefont
  {Mi\ifmmode~\check{s}\else \v{s}\fi{}ta}}\ and\ \bibinfo {author}
  {\bibfnamefont {N.}~\bibnamefont {Korolkova}},\ }\href {\doibase
  10.1103/PhysRevA.77.050302} {\bibfield  {journal} {\bibinfo  {journal} {Phys.
  Rev. A}\ }\textbf {\bibinfo {volume} {77}},\ \bibinfo {pages} {050302}
  (\bibinfo {year} {2008})}\BibitemShut {NoStop}%
\bibitem [{\citenamefont {Fedrizzi}\ \emph {et~al.}(2013)\citenamefont
  {Fedrizzi}, \citenamefont {Zuppardo}, \citenamefont {Gillett}, \citenamefont
  {Broome}, \citenamefont {Almeida}, \citenamefont {Paternostro}, \citenamefont
  {White},\ and\ \citenamefont {Paterek}}]{Fedrizzi_2013}%
  \BibitemOpen
  \bibfield  {author} {\bibinfo {author} {\bibfnamefont {A.}~\bibnamefont
  {Fedrizzi}}, \bibinfo {author} {\bibfnamefont {M.}~\bibnamefont {Zuppardo}},
  \bibinfo {author} {\bibfnamefont {G.~G.}\ \bibnamefont {Gillett}}, \bibinfo
  {author} {\bibfnamefont {M.~A.}\ \bibnamefont {Broome}}, \bibinfo {author}
  {\bibfnamefont {M.~P.}\ \bibnamefont {Almeida}}, \bibinfo {author}
  {\bibfnamefont {M.}~\bibnamefont {Paternostro}}, \bibinfo {author}
  {\bibfnamefont {A.~G.}\ \bibnamefont {White}}, \ and\ \bibinfo {author}
  {\bibfnamefont {T.}~\bibnamefont {Paterek}},\ }\href {\doibase
  10.1103/PhysRevLett.111.230504} {\bibfield  {journal} {\bibinfo  {journal}
  {Phys. Rev. Lett.}\ }\textbf {\bibinfo {volume} {111}},\ \bibinfo {pages}
  {230504} (\bibinfo {year} {2013})}\BibitemShut {NoStop}%
\bibitem [{\citenamefont {Peuntinger}\ \emph {et~al.}(2013)\citenamefont
  {Peuntinger}, \citenamefont {Chille}, \citenamefont
  {Mi\ifmmode~\check{s}\else \v{s}\fi{}ta}, \citenamefont {Korolkova},
  \citenamefont {F\"ortsch}, \citenamefont {Korger}, \citenamefont
  {Marquardt},\ and\ \citenamefont {Leuchs}}]{Peuntinger_2013}%
  \BibitemOpen
  \bibfield  {author} {\bibinfo {author} {\bibfnamefont {C.}~\bibnamefont
  {Peuntinger}}, \bibinfo {author} {\bibfnamefont {V.}~\bibnamefont {Chille}},
  \bibinfo {author} {\bibfnamefont {L.}~\bibnamefont {Mi\ifmmode~\check{s}\else
  \v{s}\fi{}ta}}, \bibinfo {author} {\bibfnamefont {N.}~\bibnamefont
  {Korolkova}}, \bibinfo {author} {\bibfnamefont {M.}~\bibnamefont
  {F\"ortsch}}, \bibinfo {author} {\bibfnamefont {J.}~\bibnamefont {Korger}},
  \bibinfo {author} {\bibfnamefont {C.}~\bibnamefont {Marquardt}}, \ and\
  \bibinfo {author} {\bibfnamefont {G.}~\bibnamefont {Leuchs}},\ }\href
  {\doibase 10.1103/PhysRevLett.111.230506} {\bibfield  {journal} {\bibinfo
  {journal} {Phys. Rev. Lett.}\ }\textbf {\bibinfo {volume} {111}},\ \bibinfo
  {pages} {230506} (\bibinfo {year} {2013})}\BibitemShut {NoStop}%
\bibitem [{\citenamefont {Vollmer}\ \emph {et~al.}(2013)\citenamefont
  {Vollmer}, \citenamefont {Schulze}, \citenamefont {Eberle}, \citenamefont
  {H\"andchen}, \citenamefont {Fiur\'a\ifmmode~\check{s}\else \v{s}\fi{}ek},\
  and\ \citenamefont {Schnabel}}]{Vollmer_2013}%
  \BibitemOpen
  \bibfield  {author} {\bibinfo {author} {\bibfnamefont {C.~E.}\ \bibnamefont
  {Vollmer}}, \bibinfo {author} {\bibfnamefont {D.}~\bibnamefont {Schulze}},
  \bibinfo {author} {\bibfnamefont {T.}~\bibnamefont {Eberle}}, \bibinfo
  {author} {\bibfnamefont {V.}~\bibnamefont {H\"andchen}}, \bibinfo {author}
  {\bibfnamefont {J.}~\bibnamefont {Fiur\'a\ifmmode~\check{s}\else
  \v{s}\fi{}ek}}, \ and\ \bibinfo {author} {\bibfnamefont {R.}~\bibnamefont
  {Schnabel}},\ }\href {\doibase 10.1103/PhysRevLett.111.230505} {\bibfield
  {journal} {\bibinfo  {journal} {Phys. Rev. Lett.}\ }\textbf {\bibinfo
  {volume} {111}},\ \bibinfo {pages} {230505} (\bibinfo {year}
  {2013})}\BibitemShut {NoStop}%
\bibitem [{\citenamefont {Eisert}\ and\ \citenamefont
  {Plenio}(2003)}]{Continuous}%
  \BibitemOpen
  \bibfield  {author} {\bibinfo {author} {\bibfnamefont {J.}~\bibnamefont
  {Eisert}}\ and\ \bibinfo {author} {\bibfnamefont {M.~B.}\ \bibnamefont
  {Plenio}},\ }\href {\doibase 10.1142/S0219749903000371} {\bibfield  {journal}
  {\bibinfo  {journal} {Int. J. Quant. Inf.}\ }\textbf {\bibinfo {volume}
  {1}},\ \bibinfo {pages} {479} (\bibinfo {year} {2003})}\BibitemShut {NoStop}%
\bibitem [{\citenamefont {Adesso}\ and\ \citenamefont
  {Illuminati}(2007)}]{Adesso_2007}%
  \BibitemOpen
  \bibfield  {author} {\bibinfo {author} {\bibfnamefont {G.}~\bibnamefont
  {Adesso}}\ and\ \bibinfo {author} {\bibfnamefont {F.}~\bibnamefont
  {Illuminati}},\ }\href {\doibase 10.1088/1751-8113/40/28/s01} {\bibfield
  {journal} {\bibinfo  {journal} {J. Phys. A}\ }\textbf {\bibinfo {volume}
  {40}},\ \bibinfo {pages} {7821} (\bibinfo {year} {2007})}\BibitemShut
  {NoStop}%
\bibitem [{\citenamefont {Weedbrook}\ \emph {et~al.}(2012)\citenamefont
  {Weedbrook}, \citenamefont {Pirandola}, \citenamefont {Garcia-Patron},
  \citenamefont {Cerf}, \citenamefont {Ralph}, \citenamefont {Shapiro},\ and\
  \citenamefont {Lloyd}}]{GaussianQuantumInfo}%
  \BibitemOpen
  \bibfield  {author} {\bibinfo {author} {\bibfnamefont {C.}~\bibnamefont
  {Weedbrook}}, \bibinfo {author} {\bibfnamefont {S.}~\bibnamefont
  {Pirandola}}, \bibinfo {author} {\bibfnamefont {R.}~\bibnamefont
  {Garcia-Patron}}, \bibinfo {author} {\bibfnamefont {N.~J.}\ \bibnamefont
  {Cerf}}, \bibinfo {author} {\bibfnamefont {T.~C.}\ \bibnamefont {Ralph}},
  \bibinfo {author} {\bibfnamefont {J.~H.}\ \bibnamefont {Shapiro}}, \ and\
  \bibinfo {author} {\bibfnamefont {S.}~\bibnamefont {Lloyd}},\ }\href
  {\doibase 10.1103/RevModPhys.84.621} {\bibfield  {journal} {\bibinfo
  {journal} {Rev. Mod. Phys.}\ }\textbf {\bibinfo {volume} {84}},\ \bibinfo
  {pages} {621} (\bibinfo {year} {2012})}\BibitemShut {NoStop}%
\bibitem [{\citenamefont {Wolf}\ \emph {et~al.}(2004)\citenamefont {Wolf},
  \citenamefont {Giedke}, \citenamefont {Kr\"uger}, \citenamefont {Werner},\
  and\ \citenamefont {Cirac}}]{PhysRevA.69.052320}%
  \BibitemOpen
  \bibfield  {author} {\bibinfo {author} {\bibfnamefont {M.~M.}\ \bibnamefont
  {Wolf}}, \bibinfo {author} {\bibfnamefont {G.}~\bibnamefont {Giedke}},
  \bibinfo {author} {\bibfnamefont {O.}~\bibnamefont {Kr\"uger}}, \bibinfo
  {author} {\bibfnamefont {R.~F.}\ \bibnamefont {Werner}}, \ and\ \bibinfo
  {author} {\bibfnamefont {J.~I.}\ \bibnamefont {Cirac}},\ }\href {\doibase
  10.1103/PhysRevA.69.052320} {\bibfield  {journal} {\bibinfo  {journal} {Phys.
  Rev. A}\ }\textbf {\bibinfo {volume} {69}},\ \bibinfo {pages} {052320}
  (\bibinfo {year} {2004})}\BibitemShut {NoStop}%
\bibitem [{\citenamefont {Vidal}\ and\ \citenamefont
  {Werner}(2002)}]{VidalNegativity}%
  \BibitemOpen
  \bibfield  {author} {\bibinfo {author} {\bibfnamefont {G.}~\bibnamefont
  {Vidal}}\ and\ \bibinfo {author} {\bibfnamefont {R.~F.}\ \bibnamefont
  {Werner}},\ }\href {\doibase 10.1103/PhysRevA.65.032314} {\bibfield
  {journal} {\bibinfo  {journal} {Phys. Rev. A}\ }\textbf {\bibinfo {volume}
  {65}},\ \bibinfo {pages} {032314} (\bibinfo {year} {2002})}\BibitemShut
  {NoStop}%
\bibitem [{\citenamefont {Plenio}(2005)}]{PlenioNegativity}%
  \BibitemOpen
  \bibfield  {author} {\bibinfo {author} {\bibfnamefont {M.~B.}\ \bibnamefont
  {Plenio}},\ }\href {\doibase 10.1103/PhysRevLett.95.090503} {\bibfield
  {journal} {\bibinfo  {journal} {Phys. Rev. Lett.}\ }\textbf {\bibinfo
  {volume} {95}},\ \bibinfo {pages} {090503} (\bibinfo {year}
  {2005})}\BibitemShut {NoStop}%
\bibitem [{\citenamefont {Eisert}(2001)}]{PhD}%
  \BibitemOpen
  \bibfield  {author} {\bibinfo {author} {\bibfnamefont {J.}~\bibnamefont
  {Eisert}},\ }\emph {\bibinfo {title} {Entanglement in quantum information
  theory}},\ \href@noop {} {Ph.D. thesis},\ \bibinfo  {school} {University of
  Potsdam} (\bibinfo {year} {2001})\BibitemShut {NoStop}%
\bibitem [{\citenamefont {Duan}\ \emph {et~al.}(2000)\citenamefont {Duan},
  \citenamefont {Giedke}, \citenamefont {Cirac},\ and\ \citenamefont
  {Zoller}}]{Duan_2000}%
  \BibitemOpen
  \bibfield  {author} {\bibinfo {author} {\bibfnamefont {L.-M.}\ \bibnamefont
  {Duan}}, \bibinfo {author} {\bibfnamefont {G.}~\bibnamefont {Giedke}},
  \bibinfo {author} {\bibfnamefont {J.~I.}\ \bibnamefont {Cirac}}, \ and\
  \bibinfo {author} {\bibfnamefont {P.}~\bibnamefont {Zoller}},\ }\href
  {\doibase 10.1103/PhysRevLett.84.2722} {\bibfield  {journal} {\bibinfo
  {journal} {Phys. Rev. Lett.}\ }\textbf {\bibinfo {volume} {84}},\ \bibinfo
  {pages} {2722} (\bibinfo {year} {2000})}\BibitemShut {NoStop}%
\bibitem [{\citenamefont {Simon}(2000)}]{Simon00}%
  \BibitemOpen
  \bibfield  {author} {\bibinfo {author} {\bibfnamefont {R.}~\bibnamefont
  {Simon}},\ }\href {\doibase 10.1103/PhysRevLett.84.2726} {\bibfield
  {journal} {\bibinfo  {journal} {Phys. Rev. Lett.}\ }\textbf {\bibinfo
  {volume} {84}},\ \bibinfo {pages} {2726} (\bibinfo {year}
  {2000})}\BibitemShut {NoStop}%
\bibitem [{\citenamefont {Hyllus}\ and\ \citenamefont
  {Eisert}(2006)}]{Hyllus_2006}%
  \BibitemOpen
  \bibfield  {author} {\bibinfo {author} {\bibfnamefont {P.}~\bibnamefont
  {Hyllus}}\ and\ \bibinfo {author} {\bibfnamefont {J.}~\bibnamefont
  {Eisert}},\ }\href {\doibase 10.1088/1367-2630/8/4/051} {\bibfield  {journal}
  {\bibinfo  {journal} {New J. Phys.}\ }\textbf {\bibinfo {volume} {8}},\
  \bibinfo {pages} {51} (\bibinfo {year} {2006})}\BibitemShut {NoStop}%
\bibitem [{\citenamefont {Eisert}\ \emph {et~al.}(2007)\citenamefont {Eisert},
  \citenamefont {Brandao},\ and\ \citenamefont {Audenaert}}]{quant-ph/0607167}%
  \BibitemOpen
  \bibfield  {author} {\bibinfo {author} {\bibfnamefont {J.}~\bibnamefont
  {Eisert}}, \bibinfo {author} {\bibfnamefont {F.~G. S.~L.}\ \bibnamefont
  {Brandao}}, \ and\ \bibinfo {author} {\bibfnamefont {K.~M.}\ \bibnamefont
  {Audenaert}},\ }\href {\doibase 10.1088/1367-2630/9/3/046} {\bibfield
  {journal} {\bibinfo  {journal} {New J. Phys.}\ }\textbf {\bibinfo {volume}
  {9}},\ \bibinfo {pages} {46} (\bibinfo {year} {2007})}\BibitemShut {NoStop}%
\bibitem [{\citenamefont {Audenaert}\ and\ \citenamefont
  {Plenio}(2006)}]{Audenaert06}%
  \BibitemOpen
  \bibfield  {author} {\bibinfo {author} {\bibfnamefont {K.~M.~R.}\
  \bibnamefont {Audenaert}}\ and\ \bibinfo {author} {\bibfnamefont {M.~B.}\
  \bibnamefont {Plenio}},\ }\href {\doibase 10.1088/1367-2630/8/11/266}
  {\bibfield  {journal} {\bibinfo  {journal} {New J. Phys.}\ }\textbf {\bibinfo
  {volume} {8}},\ \bibinfo {pages} {266} (\bibinfo {year} {2006})}\BibitemShut
  {NoStop}%
\bibitem [{\citenamefont {G\"uhne}\ \emph {et~al.}(2007)\citenamefont
  {G\"uhne}, \citenamefont {Reimpell},\ and\ \citenamefont {Werner}}]{Guehne}%
  \BibitemOpen
  \bibfield  {author} {\bibinfo {author} {\bibfnamefont {O.}~\bibnamefont
  {G\"uhne}}, \bibinfo {author} {\bibfnamefont {M.}~\bibnamefont {Reimpell}}, \
  and\ \bibinfo {author} {\bibfnamefont {R.~F.}\ \bibnamefont {Werner}},\
  }\href {\doibase 10.1103/PhysRevLett.98.110502} {\bibfield  {journal}
  {\bibinfo  {journal} {Phys. Rev. Lett.}\ }\textbf {\bibinfo {volume} {98}},\
  \bibinfo {pages} {110502} (\bibinfo {year} {2007})}\BibitemShut {NoStop}%
\bibitem [{\citenamefont {Hoelscher-Obermaier}(2017)}]{Thesis_Jason}%
  \BibitemOpen
  \bibfield  {author} {\bibinfo {author} {\bibfnamefont {J.}~\bibnamefont
  {Hoelscher-Obermaier}},\ }{\selectlanguage {english}\emph {\bibinfo {title}
  {Generation and detection of quantum entanglement in optomechanical
  systems.}}},\ \href@noop {} {Ph.D. thesis},\ \bibinfo  {school} {University
  of Vienna} (\bibinfo {year} {2017})\BibitemShut {NoStop}%
\bibitem [{\citenamefont {Rossi}\ \emph {et~al.}(2018)\citenamefont {Rossi},
  \citenamefont {Mason}, \citenamefont {Chen},\ and\ \citenamefont
  {Schliesser}}]{rossi2018observing}%
  \BibitemOpen
  \bibfield  {author} {\bibinfo {author} {\bibfnamefont {M.}~\bibnamefont
  {Rossi}}, \bibinfo {author} {\bibfnamefont {D.}~\bibnamefont {Mason}},
  \bibinfo {author} {\bibfnamefont {J.}~\bibnamefont {Chen}}, \ and\ \bibinfo
  {author} {\bibfnamefont {A.}~\bibnamefont {Schliesser}},\ }\href@noop {}
  {\enquote {\bibinfo {title} {Observing and verifying the quantum trajectory
  of a mechanical resonator},}\ } (\bibinfo {year} {2018}),\ \Eprint
  {http://arxiv.org/abs/1812.00928} {arXiv:1812.00928 [quant-ph]} \BibitemShut
  {NoStop}%
\bibitem [{\citenamefont {Gut}\ and\ \citenamefont
  {Winkler}(2019)}]{gut_corentin_2019_3901001}%
  \BibitemOpen
  \bibfield  {author} {\bibinfo {author} {\bibfnamefont {C.}~\bibnamefont
  {Gut}}\ and\ \bibinfo {author} {\bibfnamefont {K.}~\bibnamefont {Winkler}},\
  }\href {\doibase 10.5281/zenodo.3901001} {\  (\bibinfo {year} {2019}),\
  10.5281/zenodo.3901001}\BibitemShut {NoStop}%
\bibitem [{Note3()}]{Note3}%
  \BibitemOpen
  \bibinfo {note} {Ref.~\cite {Hyllus_2006} provides a \protect \textsc
  {MATLAB} function that computes the optimal witness given a covariance
  matrix.}\BibitemShut {Stop}%
\bibitem [{\citenamefont {Boyd}\ and\ \citenamefont
  {Vanderberghe}(2004)}]{Boyd2004}%
  \BibitemOpen
  \bibfield  {author} {\bibinfo {author} {\bibfnamefont {S.}~\bibnamefont
  {Boyd}}\ and\ \bibinfo {author} {\bibfnamefont {L.}~\bibnamefont
  {Vanderberghe}},\ }\href {\doibase 10.1017/CBO9780511804441} {\emph {\bibinfo
  {title} {Convex optimization}}}\ (\bibinfo  {publisher} {Cambridge University
  Press},\ \bibinfo {address} {Cambridge},\ \bibinfo {year} {2004})\BibitemShut
  {NoStop}%
\bibitem [{\citenamefont {Gross}\ \emph {et~al.}(2010)\citenamefont {Gross},
  \citenamefont {Liu}, \citenamefont {Flammia}, \citenamefont {Becker},\ and\
  \citenamefont {Eisert}}]{Compressed}%
  \BibitemOpen
  \bibfield  {author} {\bibinfo {author} {\bibfnamefont {D.}~\bibnamefont
  {Gross}}, \bibinfo {author} {\bibfnamefont {Y.-K.}\ \bibnamefont {Liu}},
  \bibinfo {author} {\bibfnamefont {S.~T.}\ \bibnamefont {Flammia}}, \bibinfo
  {author} {\bibfnamefont {S.}~\bibnamefont {Becker}}, \ and\ \bibinfo {author}
  {\bibfnamefont {J.}~\bibnamefont {Eisert}},\ }\href {\doibase
  10.1103/PhysRevLett.105.150401} {\bibfield  {journal} {\bibinfo  {journal}
  {Phys. Rev. Lett.}\ }\textbf {\bibinfo {volume} {105}},\ \bibinfo {pages}
  {150401} (\bibinfo {year} {2010})}\BibitemShut {NoStop}%
\bibitem [{\citenamefont {Groeblacher}\ \emph {et~al.}(2015)\citenamefont
  {Groeblacher}, \citenamefont {Trubarov}, \citenamefont {Prigge},
  \citenamefont {Aspelmeyer},\ and\ \citenamefont {Eisert}}]{Observation}%
  \BibitemOpen
  \bibfield  {author} {\bibinfo {author} {\bibfnamefont {S.}~\bibnamefont
  {Groeblacher}}, \bibinfo {author} {\bibfnamefont {A.}~\bibnamefont
  {Trubarov}}, \bibinfo {author} {\bibfnamefont {N.}~\bibnamefont {Prigge}},
  \bibinfo {author} {\bibfnamefont {M.}~\bibnamefont {Aspelmeyer}}, \ and\
  \bibinfo {author} {\bibfnamefont {J.}~\bibnamefont {Eisert}},\ }\href
  {\doibase 10.1038/ncomms8606} {\bibfield  {journal} {\bibinfo  {journal}
  {Nature Comm.}\ }\textbf {\bibinfo {volume} {6}},\ \bibinfo {pages} {7606}
  (\bibinfo {year} {2015})}\BibitemShut {NoStop}%
\bibitem [{\citenamefont {Titchmarsh}(1948)}]{Titchmarsh}%
  \BibitemOpen
  \bibfield  {author} {\bibinfo {author} {\bibfnamefont {E.}~\bibnamefont
  {Titchmarsh}},\ }\href@noop {} {\emph {\bibinfo {title} {Introduction to the
  theory of Fourier intergrals}}}\ (\bibinfo  {publisher} {Oxford University
  Press},\ \bibinfo {year} {1948})\BibitemShut {NoStop}%
\bibitem [{\citenamefont {Nussenzveig}(1972)}]{Nussenzveig_72}%
  \BibitemOpen
  \bibfield  {author} {\bibinfo {author} {\bibfnamefont {H.}~\bibnamefont
  {Nussenzveig}},\ }\href@noop {} {\emph {\bibinfo {title} {Causality and
  Dispersion Relations}}}\ (\bibinfo  {publisher} {Elsevier/Academic Press},\
  \bibinfo {year} {1972})\BibitemShut {NoStop}%
\bibitem [{\citenamefont {Park}\ and\ \citenamefont {Boyd}(2017)}]{qcqp}%
  \BibitemOpen
  \bibfield  {author} {\bibinfo {author} {\bibfnamefont {J.}~\bibnamefont
  {Park}}\ and\ \bibinfo {author} {\bibfnamefont {S.}~\bibnamefont {Boyd}},\
  }\href@noop {} {\enquote {\bibinfo {title} {General heuristics for nonconvex
  quadratically constrained quadratic programming},}\ } (\bibinfo {year}
  {2017}),\ \Eprint {http://arxiv.org/abs/1703.07870} {arXiv:1703.07870
  [math.OC]} \BibitemShut {NoStop}%
\bibitem [{\citenamefont {Hofer}(2015)}]{Thesis_Hofer}%
  \BibitemOpen
  \bibfield  {author} {\bibinfo {author} {\bibfnamefont {S.~G.}\ \bibnamefont
  {Hofer}},\ }{\selectlanguage {english}\emph {\bibinfo {title} {Quantum
  {control} of {optomechanical} {systems}}}},\ \href@noop {} {Ph.D. thesis},\
  \bibinfo  {school} {University of Vienna} (\bibinfo {year}
  {2015})\BibitemShut {NoStop}%
\end{thebibliography}%

\end{document}